\documentclass[aps,prl,twocolumn,superscriptaddress,longbibliography]{revtex4-2}
\usepackage{graphicx}
\usepackage{amssymb,amsmath}
\usepackage{bm}
\usepackage{dcolumn}
\usepackage{float}
\usepackage[OT1]{fontenc} 
\usepackage{url}
\usepackage{mathrsfs}
\usepackage{slashed,comment}
\usepackage{color}
\usepackage{verbatim}
\usepackage{enumitem}
\usepackage{txfonts}
\usepackage{soul,physics}
\usepackage[driverfallback=dvipdfm]{hyperref}
\hypersetup{pdfpagemode=FullScreen,colorlinks=true,breaklinks,urlcolor=blue,linkcolor=blue,citecolor=blue}

\usepackage{pifont}
\def\X#1{%
        \ding{\numexpr171+#1\relax}%
}

\usepackage{subfigure}
\usepackage{amssymb}
\usepackage{bm}
\usepackage{graphicx}
\usepackage{amsmath,amssymb}

\usepackage{tikz}
\usepackage{tikz-cd}
\usetikzlibrary{arrows}
\usetikzlibrary{intersections}
\usetikzlibrary{shapes.geometric}
\usetikzlibrary{decorations.pathmorphing, patterns,shapes}
\usetikzlibrary{decorations.markings}

\tikzset{
  mid arrow/.style={postaction={decorate,decoration={
        markings,
        mark=at position .575 with {\arrow[#1]{stealth}}
      }}},
  near arrow/.style={postaction={decorate,decoration={
        markings,
        mark=at position .375 with {\arrow[#1]{stealth}}
      }}},
   far arrow/.style={postaction={decorate,decoration={
        markings,
        mark=at position .7 with {\arrow[#1]{stealth}}
      }}},
}

\tikzset{
  baseline = -0.5ex,
  wavy/.style = {
    thick,
    decorate,
    decoration={snake,amplitude=2pt,segment length=5pt}},
  sdot/.style = {
    circle,
    draw=none,
    fill=black,
    minimum size=2.5pt,
    inner sep=0pt},
  bdot/.style = {
    circle,
    draw=none,
    fill=black,
    minimum size=4pt,
    inner sep=0pt},
  svertex/.style = {
    circle,
    draw=black,
    thick,
    fill=lightgray,
    minimum size=8pt,
    inner sep=1pt},
  mvertex/.style = {
    circle,
    draw=black,
    thick,
    fill=lightgray,
    minimum size=12pt,
    inner sep=1pt},
  bvertex/.style = {
    circle,
    draw=black,
    thick,
    fill=lightgray,
    minimum size=16pt}}

\usepackage{pdfpages} 
\usepackage{pgffor} 

\makeatletter
\AtBeginDocument{\let\LS@rot\@undefined}
\makeatother

\def\supplementfilename{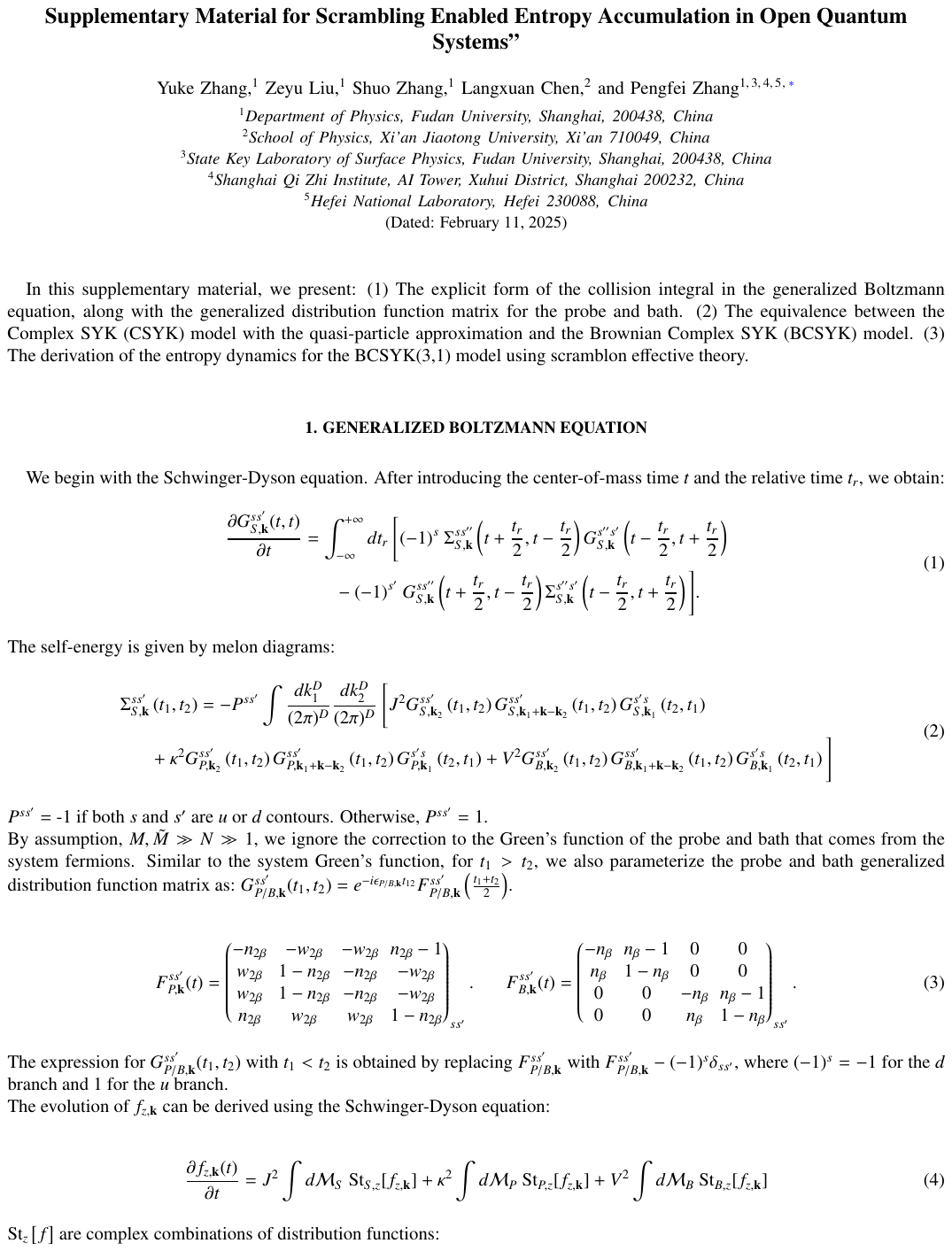}

\pdfximage{\supplementfilename}
\def\numbersupplementpages{\the\pdflastximagepages}

\newif\ifarXiv
\arXivtrue 

\begin{document}
  
  \title{Scrambling Enabled Entropy Accumulation in Open Quantum Systems}

  \author{Yuke Zhang}
  \affiliation{Department of Physics, Fudan University, Shanghai, 200438, China}
  
  \author{Zeyu Liu}
  \affiliation{Department of Physics, Fudan University, Shanghai, 200438, China}

  \author{Shuo Zhang}
  \affiliation{Department of Physics, Fudan University, Shanghai, 200438, China}

  \author{Langxuan Chen}
  \affiliation{School of Physics, Xi'an Jiaotong University, Xi'an 710049, China}

  \author{Pengfei Zhang}
  \thanks{PengfeiZhang.physics@gmail.com}
  \affiliation{Department of Physics, Fudan University, Shanghai, 200438, China}
  \affiliation{State Key Laboratory of Surface Physics, Fudan University, Shanghai, 200438, China}
  \affiliation{Shanghai Qi Zhi Institute, AI Tower, Xuhui District, Shanghai 200232, China}
  \affiliation{Hefei National Laboratory, Hefei 230088, China}
  \date{\today}

  \begin{abstract}
  In closed quantum many-body systems, initially localized information spreads throughout the system and becomes highly complex. This phenomenon, known as information scrambling, is closely related to entropy growth and quantum thermalization. Recent studies have shown that dissipation in open systems can hinder information scrambling, driving the system into a dissipative phase when the system-bath coupling is strong. However, the signature of this scrambling transition in entropy dynamics remains unexplored. In this work, we unveil a novel phenomenon in open quantum systems, termed entropy accumulation, which occurs exclusively within the scrambling phase. We consider a setup in which a probe is weakly coupled to a system that is already interacting with a bath. We calculate the increase in the second R\'enyi entropy induced by an external impulse on the system, after tracing out the probe. Despite the system-probe coupling being weak, the entropy continues to increase and eventually saturates at a finite value due to operator growth. In contrast, the entropy increase is limited by the coupling strength in the dissipative phase. The theoretical prediction is derived from both general arguments and an explicit example using generalized Boltzmann equations. Our results offer new insights into the intriguing relationship between entropy dynamics and information scrambling in open quantum systems.
  \end{abstract}
  
  \maketitle

  \emph{ \color{blue}Introduction.--} 
  In closed quantum systems, the process of quantum thermalization involves the complete scrambling of local initial conditions throughout the system after a sufficiently long unitary evolution \cite{PhysRevA.43.2046,PhysRevE.50.888}. This phenomenon, known as information scrambling \cite{Hayden:2007cs,Sekino:2008he,Shenker:2014cwa,Roberts:2014isa}, is quantified by out-of-time-ordered correlators (OTOCs) that measures the growth of operator size \cite{LaOv69,Nahum:2017yvy, Qi:2018bje, Hunter-Jones:2018otn, vonKeyserlingk:2017dyr, PhysRevX.8.031057, Dias:2021ncd, PhysRevResearch.3.L032057,  Roberts:2018aa, qi2019, Lucas:2020pgj, Lensky:2020ubw, PhysRevLett.122.216601, Chen:2019klo, Chen:2020bmq, Yin:2020oze, Zhou:2021syv, Omanakuttan:2022ikz, Ippoliti:2022vfn}. It has been extensively studied in various contexts \cite{Xu:2022vko,Bhattacharyya:2021ypq}, revealing broad implications for quantum many-body dynamics. For instance, the OTOC-RE theorem provides an exact relation between the decay of OTOCs and the increase in the second R\'enyi entropy following a quantum quench \cite{Hosur:2015ylk,Fan:2016ean,Padmanabhan:2017ekk,PhysRevResearch.1.033044,PhysRevE.102.052133,Chen:2020atj}, which has been used to experimentally measure entropy dynamics \cite{PhysRevX.7.031011}. Recent developments further extend this correspondence to the Von Neumann entropy using perturbative calculations, which involves introducing branched OTOCs for arbitrary replicas \cite{dadras2021perturbative,Zhang:2023gzl}.

  On the other hand, there is growing interest in understanding the non-equilibrium dynamics of open quantum systems. One intriguing aspect is the competition between dissipation and scrambling in the evolution of quantum information \cite{Chen:2017dbb, Zhang:2019fcy,PhysRevB.97.161114,PhysRevA.100.062106, Almheiri:2019jqq, Zhang:2023xrr, Weinstein:2022yce, PhysRevResearch.5.033085, Bhattacharya:2022gbz, Schuster:2022bot, Bhattacharjee:2022lzy, Bhattacharjee:2023uwx,PhysRevD.110.086010,Zhang:2024vsa,Zhou:2024osg}. In particular, it has been established that a transition from a scrambling phase, where the operator size exhibits exponential growth, to a dissipative phase, where the operator size decays exponentially, can be induced by tuning the system-environment coupling in both fermionic and qubit systems \cite{Zhang:2023xrr,Weinstein:2022yce,Zhang:2024vsa}. Subsequent investigations show that this transition can also be interpreted as a change in the quantum information processing capability, as revealed by the fidelity of many-body quantum teleportation protocol \cite{Zhou:2024osg}. Nevertheless, despite the close relationship between information scrambling and entropy growth, how this scrambling transition gives rise to novel phenomena in entropy dynamics remains an open question. One main challenge is that, despite the transition in operator size, the system always thermalizes for any non-zero system-bath coupling, resulting in the same saturation entropy for the system \cite{Weinstein:2022yce}.

 \begin{figure}[t]
    \centering
    \includegraphics[width=0.85\linewidth]{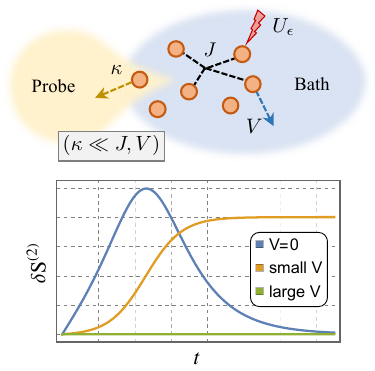}
    \caption{We present a schematic of our setup, which involves coupling a probe to an open system embedded within a bath. We compute the entropy dynamics induced by an external impulse $U_\epsilon=e^{-i\epsilon X}$ using perturbation theory in $\epsilon$. The results reveal distinct features as $V/J$ is tuned across the entanglement transition. Notably, only in the scrambling phase (for small $V$), does the entropy continuously increase, eventually saturating to a finite value, a phenomenon we refer to as entropy accumulation. }
    \label{fig:schemticas}
  \end{figure}

  In this letter, we overcome this difficulty by introducing an alternative setup. Specifically, we consider a probe weakly coupled to the open system, which is embedded within a bath, as depicted in Fig. \ref{fig:schemticas}. The total system is prepared in thermal equilibrium, and a weak external impulse is subsequently injected. After evolution, we trace out the probe and compute the entropy of the reduced density matrix. As we will demonstrate, this entropy is sensitive to the system's scrambling ability: it can increase and eventually saturate to a finite value for arbitrarily weak coupling $\kappa$ when the system is in the scrambling phase, where the perturbation remains non-trivial within the system at all times. In contrast, the entropy grows and then halts at $O(\kappa^2)$ in the dissipative phase. The behavior is also qualitatively different in a closed system, where entropy eventually decays to zero \cite{Zhang:2023gzl}. Our analysis is based on general arguments for OTOCs in open systems, supplemented by an explicit calculation using the generalized Boltzmann equation \cite{aleiner2016microscopic,Zhang:2023gzl}, which allows analytical treatment for systems with flat bands.

  \emph{ \color{blue}Setup.--} We consider an open quantum system comprising a system $S$ and a bath $B$. To be concrete, we focus on fermionic models with charge conservation, although generalizations to qubit models share the same physical picture. The system $S$ consists of multi-flavor fermion modes $c_{S,\mathbf{k}}^j$ with $j\in\{1,2,\dots, N\}$, while the bath $B$ contains modes $c_{B,\mathbf{k}}^a$, where $a \in \{1, 2, \dots, M\}$. We assume $M \gg N \gg 1$, which ensures that the Hilbert space of the bath $B$ is much larger than that of the system $S$, thereby justifying the identification of $B$ as a bath. The Hamiltonian of $S\cup B$ contains three parts $H=H_S+H_B+H_{SB}$, where the system Hamiltonian reads
  \begin{equation}
  \begin{aligned}\label{eq:HS}
  H_S=&\sum_{\mathbf{k}}\epsilon_{S,\mathbf{k}}~c_{S,\mathbf{k}}^\dagger c_{S,\mathbf{k}}^{}+\frac{1}{L }\sum_{\mathbf{k},\mathbf{k}'\mathbf{q}}\sum_{i<j,h<l}J_{ijhl}c_{S,\mathbf{k}+\mathbf{q}}^{i,\dagger} c_{S,\mathbf{k}'-\mathbf{q}}^{j,\dagger} c_{S,\mathbf{k}'}^{h}c_{S,\mathbf{k}}^{l}.
  \end{aligned}
  \end{equation}
  Here, $L$ denotes the size of the entire system. The bath consists of a series of free fermion modes $H_B=\sum_{\mathbf{k},a}\epsilon_{B,\mathbf{k}}~c_{B,\mathbf{k}}^{a,\dagger} c_{B,\mathbf{k}}^{a}$, and we assume the system-bath coupling is given by
  \begin{equation}\label{eq:HSB}
  \begin{aligned}
  H_{SB}=\frac{1}{L }&\sum_{\mathbf{k},\mathbf{k}'\mathbf{q}}\sum_{ia,b<c}V_{iabc}\left[c_{S,\mathbf{k}+\mathbf{q}}^{i,\dagger} c_{B,\mathbf{k}'-\mathbf{q}}^{a,\dagger} c_{B,\mathbf{k}'}^{b}c_{B,\mathbf{k}}^{c}+\text{H.C.}\right].
  \end{aligned}
  \end{equation}
  We assume that interaction parameters with different indices are independent Gaussian variables with zero mean and
  \begin{equation}
  \overline{(J_{ijhl})^2}={2J^2}/{N^3},\ \ \ \ \ \ \overline{(V_{ijhl})^2}={2V^2}/{M^3}.
  \end{equation} 
  The prefactor of $N$ or $M$ ensures that each system fermion $c_{S,\mathbf{k}}$ acquires a finite self-energy due to the interaction. In this model, the Heisenberg evolution induced by the system-bath coupling can couple a system operator $c_{S}$ to a bath operator $c^\dagger_Bc_Bc_B$, resulting in a reduction of the operator size within the system. Consequently, general arguments presented in \cite{Zhang:2024vsa} suggest that the system undergoes a scrambling transition from a scrambling phase to a dissipative phase as $V/J$ is increased beyond a critical value $r_c$. Alternative forms of the system-bath coupling, such as direct hopping, lead to qualitatively the same behavior.

   Next, we introduce the probe $P$, which also contains a large number of fermions $c_{P,\mathbf{k}}^m$, where $ m \in \{1, 2, \dots, \tilde{M}\}$ with $\tilde{M} \gg 1$. The Hamiltonian of the probe reads $H_P=\sum_{\mathbf{k},m}\epsilon_{P,\mathbf{k}}~c_{P,\mathbf{k}}^{m,\dagger} c_{P,\mathbf{k}}^{m}$. This choice ensures that the probe can establish extensive quantum entanglement. We consider a general coupling between the system and the probe $H_{SP}=\sum_i\kappa O_{S,i}O_{P,i}$ with system operators $O_{S,i}$ and probe operators $O_{P,i}$. Throughout the manuscript, we are focusing on the probe limit $\kappa \ll V, J$. 
   We aim to understand how the coupling to the probe affects the entropy dynamics. The setup is as follows: Initially, the total system is prepared in thermal equilibrium with $\rho_{\text{eq}}=e^{-\beta H_{\text{tot}}}/Z$, where the total Hamiltonian is $H_{\text{tot}}=H+H_P+H_{SP}$ and $Z=\text{tr}[e^{-\beta H_{\text{tot}}}]$ is the partition function \footnote{Considering an initial state without system-probe interaction leads only to a result differing by $O(\kappa^2)$.}. At time $t=0$, an external impulse is applied to the system, described by the unitary operation $U_\epsilon=e^{-i\epsilon X}$, where the Hermitian operator $X$ is supported solely in the system $S$. After evolving for a time $t$, the density matrix is given by 
   \begin{equation}
   \rho(t,\epsilon)= e^{-iH_{\text{tot}}t}~U_\epsilon~\rho_{\text{eq}}~U_\epsilon^\dagger ~e^{iH_{\text{tot}}t}.
   \end{equation}
   For a sufficiently long time $t$, thermalization predicts that the reduced density matrix of the system $S$ becomes thermal for arbitrary finite system-bath coupling $V$. To gain a more detailed understanding of entanglement properties, we instead consider the reduced density matrix $\rho_{SB}(t,\epsilon)=\text{tr}_{P}[\rho(t,\epsilon)]$ and compute the corresponding second R\'enyi entropy:
   \begin{equation}
   \begin{aligned}
   &S^{(2)}(t,\epsilon)=-\ln \text{tr}_{SB}[\rho_{SB}(t,\epsilon)^2],
   \end{aligned}
   \end{equation}
   It is straightforward to see that both system-probe coupling and a finite impulse are necessary for non-trivial dynamics. In particular, we have $S^{(2)}(t,0)=S_0$, which includes both the thermal entropy of $S\cup B$ and their entanglement entropy with the probe $P$. The full analysis of $S^{(2)}(t,\epsilon)$ is generally quite challenging. Therefore, we perform a perturbative expansion for $\epsilon \ll 1$ and focus on 
   \begin{equation}
   \delta S^{(2)}(t)\equiv \left.\frac12\frac{\partial^2}{\partial \epsilon^2} S^{(2)}(t,\epsilon)\right|_{\epsilon=0}.
   \end{equation}

   \emph{ \color{blue}General Analysis.--} For closed systems $V=0$, $\delta S^{(2)}(t)$ has been computed in \cite{dadras2021perturbative,Zhang:2023gzl}, revealing a close analogy to the Page curve \cite{Page:1993df} (see the blue curve in FIG. \ref{fig:schemticas}). At short times, the entropy increases since excitations induced by $U_\epsilon$ entangles the system and the probe. On the other hand, the perturbed entropy vanishes in the long-time limit due to the thermalization of the system $S$. After a finite coupling to the bath is turned on, a naive argument already suggests that the results will differ qualitatively: Since $\kappa \ll V$, the excitations in the system predominantly relax into the bath, and the coupling to the probe $P$ does not lead to an increase in entropy. As we will demonstrate, this picture is only valid in the dissipative phase. In the scrambling phase, entropy continues to accumulate due to the growth of operator size within the system $S$.
  
   We begin with a expansion of the density matrix 
   \begin{equation}
   \rho_{SB}(t,\epsilon)= \rho_{0}+ \delta\rho_{1} \epsilon+ \delta\rho_{2} \epsilon^2+O(\epsilon^3).
   \end{equation}
   Here, $\rho_{0}=\text{tr}_P[\rho_{\text{eq}}]$ is the unperturbed reduced density matrix. $\delta\rho_1= -i \text{tr}_P[e^{-iH_{\text{tot}}t}~[X,\rho_0] ~e^{iH_{\text{tot}}t}]$ contains a single insertion of the operator $X$, while $\delta\rho_2=-\frac12\text{tr}_P[e^{-iH_{\text{tot}}t}~[X,[X,\rho_0]] ~e^{iH_{\text{tot}}t}]$ contains two insertions. Then, we identify two types of contributions to $\delta S^{(2)}(t)$:
   \begin{equation}\label{eq:x1x2}
   \delta S^{(2)}(t)=\underbrace{-\frac{2\text{tr}_{SB}[\rho_0\delta\rho_2]}{\text{tr}_{SB}[\rho_0^2]}}_{\text{\X1}(t)}\underbrace{-\frac{\text{tr}_{SB}[\delta\rho_1^2]}{\text{tr}_{SB}[\rho_0^2]}}_{\text{\X2}(t)}.
   \end{equation}
   Their time evolutions are governed by different physical processes.
   Let us first consider \X1, where a typical contribution is presented pictorially using path-integral formalism:
   \begin{equation}\label{eq:contribution1}
   \begin{aligned}
   \text{\X1}(t)&\supseteq -2 \times
   \begin{tikzpicture}[scale = 0.7,baseline={([yshift=-3.2pt]current bounding box.center)}]
   \draw[blue,thick] (-0.6,0) arc(180:8:0.6 and 0.6);
   \draw[blue,thick] (-0.6,0) arc(180:352:0.6 and 0.6);
   \draw[black,thick] (-0.8,0) arc(180:15:0.8 and 0.8);
   \draw[black,thick] (-0.8,0) arc(180:345:0.8 and 0.8);
      \draw[blue] (-0.25,0.25) node{\scriptsize$P$};
      \draw[black] (-0.725,0.75) node{\scriptsize$SB$};
   \draw[black,thick,near arrow] (0.77,0.2)-- (2.2,0.2);
   \draw[black,thick,far arrow] (2.2,-0.2)--(0.77,-0.2);
   \draw[blue,thick] (0.58,0.075)-- (1.9,0.075);
   \draw[blue,thick] (1.9,-0.075)-- (0.58,-0.075);  
   \draw[blue,thick] (1.9,0.075) arc(90:-90:0.075 and 0.075);

   \draw[blue,thick] (2.37,0) arc(180:-180:0.6 and 0.6);
   \draw[black,thick] (2.2,0.2) arc(165:-165:0.8 and 0.8);

      \draw[black,gray] (3,0.) node{\scriptsize$\rho_0$};
      \draw[gray,dashed,thick] (2.1,0.9)-- (2.1,-0.9);
      \draw[gray,dashed,thick] (3.9,0.9)-- (3.9,-0.9);
      \draw[gray,dashed,thick] (3.9,0.9)-- (2.1,0.9);
      \draw[gray,dashed,thick] (3.9,-0.9)-- (2.1,-0.9);

   \filldraw[red] (0.8,0.2) circle (1.2pt) node {$ $};
    \filldraw[red] (0.8,-0.2) circle (1.2pt) node {$ $};

      \draw[black,dashed] (1.6,0.2)-- (1.6,0.075);
  \filldraw[red] (1.6,0.2) circle (1.2pt) node {$ $};
    \filldraw[blue] (1.6,0.075) circle (1.2pt) node {$ $};

      \draw[black,dashed] (1.8,-0.2)-- (1.8,-0.075);
  \filldraw[red] (1.8,-0.2) circle (1.2pt) node {$ $};
    \filldraw[blue] (1.8,-0.075) circle (1.2pt) node {$ $};

   \filldraw (1,0.2) circle (0pt) node[above] {\scriptsize$X$};
   \filldraw (1,-0.2) circle (0pt) node[below] {\scriptsize$X$};
\end{tikzpicture}\\&\sim -2~\Big(G^W_{2\beta}+\kappa^2 G_P*F_{\text{NO}}\Big),
\end{aligned}
   \end{equation}
   The black/blue solid lines are the path-integral contour for the system embedded in the bath or the probe, respectively. The arcs represent imaginary-time evolution, while the straight lines correspond to real-time evolution. Dots indicate the insertion of operators. Here, $G^W_{2\beta}=\left.{\text{tr}_{SB}[\rho_0 X \rho_0 X]}/{\text{tr}_{SB}[\rho_0^2]}\right|_{\kappa=0}$ and we retain terms up to order $O(\kappa^2)$. For conciseness, we use $G_P * F_{\text{NO}}$ to denote the contribution from the probe two-point function $G_P(\theta_1,\theta_2) = \langle O_{P,i}(\theta_1) O_{P,i}(\theta_2) \rangle$ and the normal-ordered four-point function $F_{\text{NO}}(\theta_1,\theta_2) = \langle X(-i\beta) O_{S,i}(\theta_1) O_{S,i}(\theta_2) X(0) \rangle_c$ on the path-integral contour, integrated over the complex time variables $\theta_j=t_j-i\tau_j$ with $t_j\in[0,t]$ and $\tau_j\in[0,2\beta)$. In this case, the normal four-point function decays as $t_1$ and $t_2$ increases, and the expansion in \eqref{eq:contribution1} holds for arbitrary time $t$. Therefore, in the limit of $\kappa \ll V, J$, we can safely take $\text{\X1}(t) = \text{\X1}(0)$ as a constant for arbitrary system-bath couplings $V/J$.

   As a comparison, in the calculation of \X2, both replicas involve forward and backward evolutions. A typical contribution not only includes the normal four-point function, but also OTOCs: 
   \begin{equation}\label{eq:contribution2}
   \begin{aligned}
   \text{\X2}(t)\supseteq
   \begin{tikzpicture}[scale = 0.7,baseline={([yshift=-3.2pt]current bounding box.center)}]
   \draw[blue,thick] (-0.6,0) arc(180:8:0.6 and 0.6);
   \draw[blue,thick] (-0.6,0) arc(180:352:0.6 and 0.6);
   \draw[black,thick] (-0.8,0) arc(180:15:0.8 and 0.8);
   \draw[black,thick] (-0.8,0) arc(180:345:0.8 and 0.8);
      \draw[blue] (-0.25,0.25) node{\scriptsize$P$};
      \draw[black] (-0.725,0.75) node{\scriptsize$SB$};
   \draw[black,thick] (0.77,0.2)-- (2,0.2);
   \draw[black,thick] (2,-0.2)--(0.77,-0.2);
   \draw[blue,thick] (0.58,0.075)-- (1.9,0.075);
   \draw[blue,thick] (1.9,-0.075)-- (0.58,-0.075);  
      
   \draw[blue,thick] (1.9,0.075) arc(90:-90:0.075 and 0.075);

      \draw[blue,thick] (-0.6,-1.8) arc(180:8:0.6 and 0.6);
   \draw[blue,thick] (-0.6,-1.8) arc(180:352:0.6 and 0.6);
   \draw[black,thick] (-0.8,-1.8) arc(180:15:0.8 and 0.8);
   \draw[black,thick] (-0.8,-1.8) arc(180:345:0.8 and 0.8);
      \draw[blue] (-0.25,-1.55) node{\scriptsize$P$};
      \draw[black] (-0.725,-1.05) node{\scriptsize$SB$};
   \draw[black,thick] (0.77,-1.6)-- (2,-1.6);
   \draw[black,thick] (2,-2)--(0.77,-2);
   \draw[blue,thick] (0.58,-1.725)-- (1.9,-1.725);
   \draw[blue,thick] (1.9,-1.875)-- (0.58,-1.875);  

   \draw[black,thick] (1.98,-0.2)-- (1.98,-1.6);
    \draw[black,thick] (1.98,0.2)-- (1.98,0.6);
    \draw[black,thick] (1.98,-2)-- (1.98,-2.4);

    \draw[black,thick] (1.92,0.4)-- (2.04,0.4);
    \draw[black,thick] (1.92,-2.2)-- (2.04,-2.2);
      
   \draw[blue,thick] (1.9,-1.725) arc(90:-90:0.075 and 0.075);

   \filldraw[red] (0.8,0.2) circle (1.2pt) node {$ $};

      \filldraw[red] (0.8,-1.6) circle (1.2pt) node {$ $};

   \filldraw (1,0.2) circle (0pt) node[above] {\scriptsize$X$};
   \filldraw (1,-1.6) circle (0pt) node[above] {\scriptsize$X$};

  \draw[black,dashed] (1.6,0.2)-- (1.6,0.075);
  \filldraw[red] (1.6,0.2) circle (1.2pt) node {$ $};
    \filldraw[blue] (1.6,0.075) circle (1.2pt) node {$ $};

      \draw[black,dashed] (1.8,-0.2)-- (1.8,-0.075);
  \filldraw[red] (1.8,-0.2) circle (1.2pt) node {$ $};
    \filldraw[blue] (1.8,-0.075) circle (1.2pt) node {$ $};

\end{tikzpicture}
   \sim G^W_{2\beta}+\kappa^2 G_P*F_{\text{OTO}}.
\end{aligned}
   \end{equation}
   Again, we keep the result to order $O(\kappa^2)$ and neglect the contribution from the normal four-point function. Here, $F_{\text{OTO}}(\theta_1, \theta_2) = \pm\langle X(-i\beta) O_{S,i}(\theta_1) X(0) O_{S,i}(\theta_2) \rangle_c$, where negative sign is chosen when both $X$ and $O_S$ are fermionic. The presence of OTOCs leads to an important distinction in the entropy dynamics between the dissipative and scrambling phases. In the dissipative phase, where $V/J > r_c$, the operator rapidly leaks into the bath \cite{Zhang:2023xrr, Weinstein:2022yce, Zhang:2024vsa}. Consequently, the OTOC, which measures the operator size, decays exponentially as $t_1$ and $t_2$ increase, similar to the normal four-point correlator. In this case, we can also approximate $\text{\X2}(t) = \text{\X2}(0) = -\text{\X1}(0)$ as a constant, and $\delta S^{(2)}(t) = 0$ up to small corrections of order $O(\kappa^2)$. 

   In the scrambling phase with $V/J<r_c$, the operator size, and thus the connected part of the OTOC, grows exponentially as $e^{\varkappa t}$. Here, $\varkappa$ is the quantum Lyapunov exponent \cite{Shenker:2014cwa,Roberts:2014isa,maldacena2016bound}, including contributions from both the system's intrinsic scrambling and bath-driven dissipation. Therefore, for arbitrarily small $\kappa$, the contribution from the second term ultimately becomes significant for $t \sim \varkappa^{-1}|\ln (\kappa/J)|$. By performing a resummation over all higher-order contributions, we generally expect \cite{gu2022two,Stanford:2021bhl,sizenewpaper}
   \begin{equation}
    \text{\X2}(t)=-\text{\X1}(0)\times g(\kappa^2 e^{\varkappa t}/J^2).
   \end{equation}
   Here, $g(x)$ is a scaling function, which decays monotonically from $1$ to a finite value as $x$ increases. This expectation will be confirmed by the example presented later sections. Therefore, we have $\delta S^{(2)}(t) = \text{\X1}(0)(1-g(\kappa^2 e^{\varkappa t}/J^2))$, showing an accumulation of entropy for an extended time window, as illustrated in FIG. \ref{fig:schemticas}.  
   Physically, this is because, despite the system rapidly thermalizing due to its entanglement with the bath, the initial perturbation induced by the operator $X$ remains non-trivially supported within the system during the scrambling phase. Therefore, adding a coupling between the system and the probe can continuously establish entanglement. More quantitatively, the increase in entropy is proportional to the number of qubits affected by the perturbation, which is described by the operator size. This leads to the exponential growth of entropy at early times.

   \emph{ \color{blue}Generalized Boltzmann Equation.--} Now, we provide a explicit analysis of $  \delta S^{(2)}(t)$, assuming the system can be described by quasi-particles with weak interactions. We make an explicit choice for the system-probe coupling 
   \begin{equation}
   H_{SP}= \frac{1}{L }\sum_{\mathbf{k},\mathbf{k}'\mathbf{q}}\sum_{im,n<p}\kappa_{imnp}\left[c_{S,\mathbf{k}+\mathbf{q}}^{i,\dagger} c_{P,\mathbf{k}'-\mathbf{q}}^{m,\dagger} c_{P,\mathbf{k}'}^{n}c_{P,\mathbf{k}}^{p}+\text{H.C.}\right],
   \end{equation}
   with random couplings $\overline{(\kappa_{imnp})^2}=2\kappa^2/\tilde{M}^3$. We further fix the perturbation operator $X=c_{S,-\mathbf{k}}^1+c_{S,\mathbf{k}}^{1,\dagger} $ \footnote{Equivalently, we can choose $X = i\eta(c_{S,-\mathbf{k}}^1+c_{S,\mathbf{k}}^{1,\dagger})$, with the auxiliary Majorana fermion mode $\eta$, to make the perturbation operator $X$ bosonic. }. In this case, both $\text{\X1}(t)$ and $\text{\X2}(t)$ are determined by two-point functions on the entropy contour \cite{Zhang:2023gzl}:
   \begin{equation}\label{eq:contour}
\begin{tikzpicture}[scale = 0.75,baseline={([yshift=-3.2pt]current bounding box.center)}]
   \draw[blue,thick] (-0.6,0) arc(180:8:0.6 and 0.6);
   \draw[blue,thick] (-0.6,0) arc(180:352:0.6 and 0.6);
   \draw[black,thick] (-0.8,0) arc(180:15:0.8 and 0.8);
   \draw[black,thick] (-0.8,0) arc(180:345:0.8 and 0.8);
      \draw[blue] (-0.25,0.25) node{\scriptsize$P$};
      \draw[black] (-0.725,0.75) node{\scriptsize$SB$};
   \draw[black,thick,mid arrow] (0.77,0.2)-- (3,0.2);
   \draw[black,thick,mid arrow] (3,-0.2)--(0.77,-0.2);
   \draw[blue,thick] (0.58,0.075)-- (2.9,0.075);
   \draw[blue,thick] (2.9,-0.075)-- (0.58,-0.075);  
      
   \draw[blue,thick] (2.9,0.075) arc(90:-90:0.075 and 0.075);

      \draw[blue,thick] (-0.6,-1.8) arc(180:8:0.6 and 0.6);
   \draw[blue,thick] (-0.6,-1.8) arc(180:352:0.6 and 0.6);
   \draw[black,thick] (-0.8,-1.8) arc(180:15:0.8 and 0.8);
   \draw[black,thick] (-0.8,-1.8) arc(180:345:0.8 and 0.8);
      \draw[blue] (-0.25,-1.55) node{\scriptsize$P$};
      \draw[black] (-0.725,-1.05) node{\scriptsize$SB$};
   \draw[black,thick,mid arrow] (0.77,-1.6)-- (3,-1.6);
   \draw[black,thick,mid arrow] (3,-2)--(0.77,-2);
   \draw[blue,thick] (0.58,-1.725)-- (2.9,-1.725);
   \draw[blue,thick] (2.9,-1.875)-- (0.58,-1.875);  

   \draw[black,thick] (2.98,-0.2)-- (2.98,-1.6);
    \draw[black,thick] (2.98,0.2)-- (2.98,0.6);
    \draw[black,thick] (2.98,-2)-- (2.98,-2.4);

    \draw[black,thick] (2.92,0.4)-- (3.04,0.4);
    \draw[black,thick] (2.92,-2.2)-- (3.04,-2.2);
      
   \draw[blue,thick] (2.9,-1.725) arc(90:-90:0.075 and 0.075);

  \draw[red] (1.9,-0.5) node{\scriptsize$d_2$};
  \draw[red] (1.9,0.5) node{\scriptsize$u_2$};

  \draw[red] (1.9,-2.3) node{\scriptsize$d_1$};
  \draw[red] (1.9,-1.3) node{\scriptsize$u_1$};

    \draw[black] (0.93,0.5) node{\scriptsize$-\infty$};
    \draw[black] (2.8,0.5) node{\scriptsize$0$};
\end{tikzpicture}
   \end{equation}

 \begin{figure*}[t]
    \centering
    \includegraphics[width=0.95\linewidth]{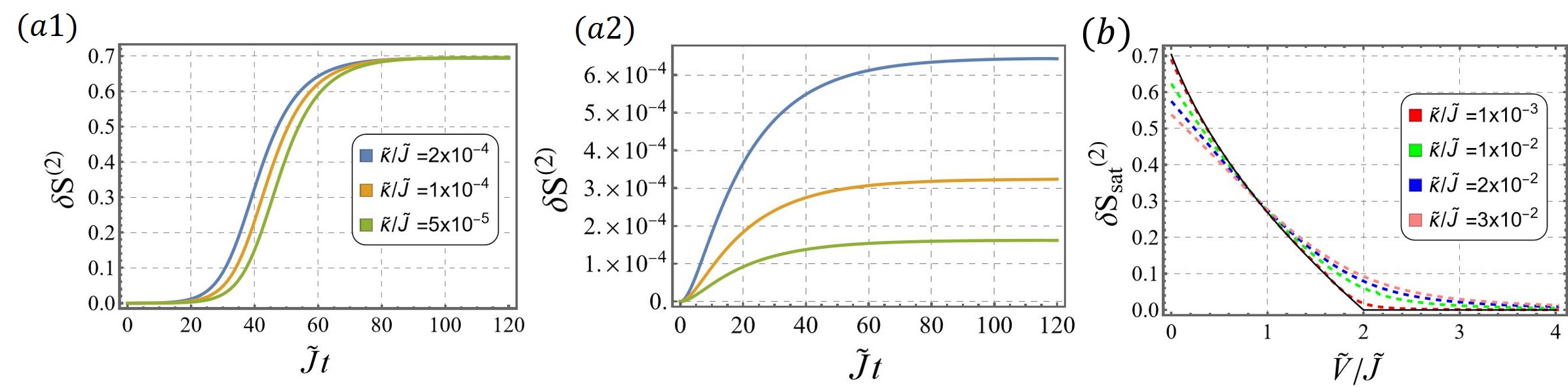}
    \caption{We present the numerical results of entropy dynamics obtained by solving the Boltzmann equation \eqref{eq:boltz}. (a1). The evolution of entropy for $\tilde{V}/\tilde{J}=0.01$ in the scrambling phase. (a2). The evolution of entropy for $\tilde{V}/\tilde{J}=2.5$ in the dissipative phase. (b). The saturation entropy as a function of $\tilde{V}/\tilde{J}$, for various $\tilde{\kappa} /\tilde{J}$. Here, the black solid line represents the prediction from the analytical solutions as $\tilde{\kappa} \rightarrow 0$.}
    \label{fig:num}
  \end{figure*}

   Compared to \eqref{eq:contribution2}, we extend the real-time evolution to $-\infty$, and the imaginary-time evolution becomes a boundary condition, similar to the conventional Keldysh contour \cite{kamenev2011field}. We label the four branches of the contour as forward (u) / backward (d) and world 1 / world 2. The contour reduces to \eqref{eq:contribution1} when all operators are inserted at finite times within a single world, as a consequence of unitarity. Our convention is to represent fields, including $c_{S,\mathbf{k}}^{i,s}$, $c_{B,\mathbf{k}}^{a,s}$, and $c_{P,\mathbf{k}}^{m,s}$, as vectors in the branch space with the ordering $s\in(d_1, u_1, d_2, u_2)$. To the leading order in the semi-classical expansion, the traditional Boltzmann equation assumes that the interaction only alters the distribution of particles, without affecting the dispersion relations \cite{kamenev2011field}. Following this spirit, we take $G^{ss'}_{S,\mathbf{k}}(t_1,t_2)=\langle c_{S,\mathbf{k}}^{i,s}(t_1)c_{S,\mathbf{k}}^{i,s',\dagger}(t_2)\rangle = e^{-i\epsilon_{S,\mathbf{k}}t_{12}}F^{ss'}_{S,\mathbf{k}}(\frac{t_1+t_2}2)$ for $t_1>t_2$, where we introduced the generalized distribution function matrix \cite{Zhang:2023gzl}
   \begin{equation}
  F_{S,\mathbf{k}}^{ss'}(t)=
  \begin{pmatrix}
-f_{0,\mathbf{k}}(t)&-f_{2,\mathbf{k}}(t)&-f_{3,\mathbf{k}}(t)&-f_{5,\mathbf{k}}(t)\\
f_{1,\mathbf{k}}(t)&1-f_{0,\mathbf{k}}(t)&-f_{4,\mathbf{k}}(t)&-f_{3,\mathbf{k}}(t)\\
f_{3,\mathbf{k}}(t)&f_{5,\mathbf{k}}(t)&-f_{0,\mathbf{k}}(t)&-f_{2,\mathbf{k}}(t)\\
f_{4,\mathbf{k}}(t)&f_{3,\mathbf{k}}(t)&f_{1,\mathbf{k}}(t)&1-f_{0,\mathbf{k}}(t)
\end{pmatrix}_{ss'}.
   \end{equation}
  The expression for $G^{ss'}_{S,\mathbf{k}}(t_1,t_2)$ with $t_1<t_2$ is given by replacing $F^{ss'}_{S,\mathbf{k}}$ by $F^{ss'}_{S,\mathbf{k}}-(-1)^{s}\delta_{ss'}$, where $(-1)^s = -1$ for the $d$ branch and $1$ for the $u$ branch. This ensures the correct commutation relation for fermion operators. The initial condition at $t=0$ can be determined by neglecting all interactions, which gives 
  \begin{equation}
  \begin{aligned}
  &f_{1,\mathbf{k}}(0)=f_{2,\mathbf{k}}(0)=f_{3,\mathbf{k}}(0)=w_{2\beta,\mathbf{k}},\\
  &f_{0,\mathbf{k}}(0)=f_{4,\mathbf{k}}(0)=f_{5,\mathbf{k}}(0)=n_{2\beta,\mathbf{k}},
  \end{aligned}
  \end{equation}
  where we have introduced $w_{2\beta,\mathbf{k}}=[2\cosh (\beta(\epsilon_{S,\mathbf{k}}-\mu))]^{-1}$ and $n_{2\beta,\mathbf{k}}=[{e^{2\beta (\epsilon_S(k)-\mu)}+1}]^{-1}$. The evolution of $f_{z,\mathbf{k}}(t)$ with $z\in\{0,1,...,5\}$ is derived by using the Schwinger-Dyson equation of the two-point function, which takes the form of generalized Boltzmann equation:
  \begin{equation}\label{eq:boltz}
  \begin{aligned}
  \frac{\partial f_{z,\mathbf{k}}(t)}{\partial t}&=J^2\int d\mathcal{M}_S~\text{St}_{S,z}[f_{z,\mathbf{k}}]+V^2\int d\mathcal{M}_B~\text{St}_{B,z}[f_{z,\mathbf{k}}]\\
  &+\kappa^2\int d\mathcal{M}_P~\text{St}_{P,z}[f_{z,\mathbf{k}}].
  \end{aligned}
  \end{equation}
  Here, $d\mathcal{M}_{S/B/P}$ represents the phase space integral for the contribution from scattering processes involving four system fermions, or one system fermion and three bath/probe fermions. $\text{St}_{S/B/P,z}[f_{z,\mathbf{k}}]$ represents complex combinations of distribution functions, where a detailed expression is provided in the supplementary material \cite{sm}. By evolving the Boltzmann equation backward in time, Eq. \eqref{eq:x1x2} gives a direct relation between $\delta S^{(2)}(t)$ and the distribution functions
  \begin{equation}
  \begin{aligned}
  \delta S^{(2)}(t)=&2-2(f_{1,\mathbf{k}}(-t)+f_{2,\mathbf{k}}(-t))\\&-2(f_{4,\mathbf{k}}(-t)+f_{5,\mathbf{k}}(-t)-2f_{3,\mathbf{k}}(-t)).
  \end{aligned}
  \end{equation}
  
  While the Boltzmann equation is valid for arbitrary dispersion relations, we now focus on the special flat-band limit, where $\epsilon_{S/B/P,\mathbf{k}} = \epsilon$. In this case, analytical solutions are available in the limit of $\kappa \rightarrow 0$. From now on, we neglect the momentum indices and introduce the effective coupling 
  \begin{equation}
  J^2\int d\mathcal{M}_S=\tilde{J}, \ \ \ V^2\int d\mathcal{M}_B=\tilde{V},\ \ \ \kappa^2\int d\mathcal{M}_P=\tilde{\kappa}.
  \end{equation}
  We first present numerical results to demonstrate the differences in entropy dynamics across the scrambling transition in various phases. In FIG. \ref{fig:num} (a1-a2), we show two representative results by choosing $\tilde{V}/\tilde{J}=0.01$ in the scrambling phase and $\tilde{V}/\tilde{J}=2.5$ in the dissipative phase. The result shows that, in the scrambling phase, the entropy grows and saturates to a finite value, while a decrease in $\tilde{\kappa}$ only delays the saturation time. In contrast, in the dissipative phase, the long-time plateau of the entropy is proportional to $\tilde{\kappa}$ and therefore vanishes as $\tilde{\kappa} \rightarrow 0$. We further plot the saturation entropy $\delta S^{(2)}(\infty)$ as a function of $\tilde{V}/\tilde{J}$ for different values of $\tilde{\kappa}/\tilde{J}$, which clearly reveals a qualitative difference depending on the scrambling ability, consistent with our general analysis. 
  
  We can also derive the theoretical prediction of saturation entropy in the limit of $\tilde{\kappa} \rightarrow 0$: We approximate that all single-replica distribution functions ($f_0$, $f_1$, and $f_2$) remain at their initial values. Moreover, the Boltzmann equation for other distributions collapses after identifying $f_4(t)=n_{2\beta}f_3(t)/w_{2\beta}$ and $f_5(t)=(1-n_{2\beta})f_3(t)/w_{2\beta}$, which gives 
  \begin{equation}
\frac{d f_3(t)}{dt}=w_{2\beta}^3 \tilde{V}-w_{2\beta}^2\left(\tilde{V}+\tilde{J}\right)f_3(t)+\tilde{J}f_3(t)^3.
  \end{equation}
  We first investigate perturbations near the initial value $f_3(0) = w_{2\beta}$. Introducing $f_3(t) = f_3(0) + \delta f_3(t)$ and retaining terms up to linear order, we obtain the equation 
  \begin{equation}
\frac{d \delta f_3}{dt} = n_{2\beta}(1-n_{2\beta}) (2\tilde{J} - \tilde{V}) \delta f_3.
\end{equation}
  This shows that when $\tilde{V} > 2\tilde{J}$, the perturbation induced by the system-probe coupling decays exponentially, so we can approximate $f_3(t) \approx f_3(0)$ and $\delta S^{(2)}(\infty)=0$. This corresponds to the dissipative phase. On the other hand, for $\tilde{V} < 2\tilde{J}$, $f_3 = w_{2\beta}$ becomes an unstable fixed point with a Lyapunov exponent $\varkappa = n_{2\beta}(1-n_{2\beta}) (2\tilde{J} - \tilde{V})$. The initial perturbation is amplified over time, and the long-time solution is described by the stable fixed point at $f_3 = w_{2\beta} \left( \sqrt{1 + 4 \frac{\tilde{V}}{\tilde{J}}} - 1 \right)$. This leads to the saturated entropy:
  \begin{equation}
  \delta S^{(2)}(\infty)= \theta(2\tilde{J}-\tilde{V})~\left(1-2{\sqrt{n_{2\beta}(1-n_{2\beta})}}\right)\left(3-\sqrt{1+4{\tilde{V}}/{\tilde{J}}}\right).
  \end{equation}
  Here, $\theta(x)$ is the Heaviside step function. This theoretical prediction is also plotted in Fig. \ref{fig:num} using a black solid line, which accurately captures the limit of $\tilde{\kappa}/\tilde{J} \rightarrow 0$.

  \emph{ \color{blue}Summary.--} In this work, we study the entropy dynamics induced by an external impulse in an open quantum system coupled to a probe. We find that the evolution is strongly dependent on the system's scrambling ability. When the system is in the scrambling phase, the entropy increases and saturates to a finite value for arbitrarily small system-probe coupling, as the perturbation remains non-trivially supported within the system due to operator growth. In contrast, in the dissipative phase, the entropy increase vanishes as $\kappa \rightarrow 0$. The results are derived using both general arguments and explicit calculations based on the generalized Boltzmann equation. We expect these findings to be readily testable in state-of-the-art experimental platforms \cite{Islam:2015mom,doi:10.1126/science.aaf6725,Brydges:2019wut,Elben:2020hpu,2012PhRvL.108k0503V,2018PhRvL.120e0406E,2019PhRvA..99e2323E}.

  \textit{Acknowledgments.}
  We thank Chengshu Li for helpful discussions. This project is supported by the Shanghai Rising-Star Program under grant number 24QA2700300, the NSFC under grant 12374477, and the Innovation Program for Quantum Science and Technology ZD0220240101.

\bibliography{main.bbl}

\begin{thebibliography}{68}%
\makeatletter
\providecommand \@ifxundefined [1]{%
 \@ifx{#1\undefined}
}%
\providecommand \@ifnum [1]{%
 \ifnum #1\expandafter \@firstoftwo
 \else \expandafter \@secondoftwo
 \fi
}%
\providecommand \@ifx [1]{%
 \ifx #1\expandafter \@firstoftwo
 \else \expandafter \@secondoftwo
 \fi
}%
\providecommand \natexlab [1]{#1}%
\providecommand \enquote  [1]{``#1''}%
\providecommand \bibnamefont  [1]{#1}%
\providecommand \bibfnamefont [1]{#1}%
\providecommand \citenamefont [1]{#1}%
\providecommand \href@noop [0]{\@secondoftwo}%
\providecommand \href [0]{\begingroup \@sanitize@url \@href}%
\providecommand \@href[1]{\@@startlink{#1}\@@href}%
\providecommand \@@href[1]{\endgroup#1\@@endlink}%
\providecommand \@sanitize@url [0]{\catcode `\\12\catcode `\$12\catcode
  `\&12\catcode `\#12\catcode `\^12\catcode `\_12\catcode `\%12\relax}%
\providecommand \@@startlink[1]{}%
\providecommand \@@endlink[0]{}%
\providecommand \url  [0]{\begingroup\@sanitize@url \@url }%
\providecommand \@url [1]{\endgroup\@href {#1}{\urlprefix }}%
\providecommand \urlprefix  [0]{URL }%
\providecommand \Eprint [0]{\href }%
\providecommand \doibase [0]{https://doi.org/}%
\providecommand \selectlanguage [0]{\@gobble}%
\providecommand \bibinfo  [0]{\@secondoftwo}%
\providecommand \bibfield  [0]{\@secondoftwo}%
\providecommand \translation [1]{[#1]}%
\providecommand \BibitemOpen [0]{}%
\providecommand \bibitemStop [0]{}%
\providecommand \bibitemNoStop [0]{.\EOS\space}%
\providecommand \EOS [0]{\spacefactor3000\relax}%
\providecommand \BibitemShut  [1]{\csname bibitem#1\endcsname}%
\let\auto@bib@innerbib\@empty
\bibitem [{\citenamefont {Deutsch}(1991)}]{PhysRevA.43.2046}%
  \BibitemOpen
  \bibfield  {author} {\bibinfo {author} {\bibfnamefont {J.~M.}\ \bibnamefont
  {Deutsch}},\ }\bibfield  {title} {\bibinfo {title} {Quantum statistical
  mechanics in a closed system},\ }\href
  {https://doi.org/10.1103/PhysRevA.43.2046} {\bibfield  {journal} {\bibinfo
  {journal} {Phys. Rev. A}\ }\textbf {\bibinfo {volume} {43}},\ \bibinfo
  {pages} {2046} (\bibinfo {year} {1991})}\BibitemShut {NoStop}%
\bibitem [{\citenamefont {Srednicki}(1994)}]{PhysRevE.50.888}%
  \BibitemOpen
  \bibfield  {author} {\bibinfo {author} {\bibfnamefont {M.}~\bibnamefont
  {Srednicki}},\ }\bibfield  {title} {\bibinfo {title} {Chaos and quantum
  thermalization},\ }\href {https://doi.org/10.1103/PhysRevE.50.888} {\bibfield
   {journal} {\bibinfo  {journal} {Phys. Rev. E}\ }\textbf {\bibinfo {volume}
  {50}},\ \bibinfo {pages} {888} (\bibinfo {year} {1994})}\BibitemShut
  {NoStop}%
\bibitem [{\citenamefont {Hayden}\ and\ \citenamefont
  {Preskill}(2007)}]{Hayden:2007cs}%
  \BibitemOpen
  \bibfield  {author} {\bibinfo {author} {\bibfnamefont {P.}~\bibnamefont
  {Hayden}}\ and\ \bibinfo {author} {\bibfnamefont {J.}~\bibnamefont
  {Preskill}},\ }\bibfield  {title} {\bibinfo {title} {{Black holes as mirrors:
  Quantum information in random subsystems}},\ }\href
  {https://doi.org/10.1088/1126-6708/2007/09/120} {\bibfield  {journal}
  {\bibinfo  {journal} {JHEP}\ }\textbf {\bibinfo {volume} {09}},\ \bibinfo
  {pages} {120}},\ \Eprint {https://arxiv.org/abs/0708.4025} {arXiv:0708.4025
  [hep-th]} \BibitemShut {NoStop}%
\bibitem [{\citenamefont {Sekino}\ and\ \citenamefont
  {Susskind}(2008)}]{Sekino:2008he}%
  \BibitemOpen
  \bibfield  {author} {\bibinfo {author} {\bibfnamefont {Y.}~\bibnamefont
  {Sekino}}\ and\ \bibinfo {author} {\bibfnamefont {L.}~\bibnamefont
  {Susskind}},\ }\bibfield  {title} {\bibinfo {title} {{Fast Scramblers}},\
  }\href {https://doi.org/10.1088/1126-6708/2008/10/065} {\bibfield  {journal}
  {\bibinfo  {journal} {JHEP}\ }\textbf {\bibinfo {volume} {10}},\ \bibinfo
  {pages} {065}},\ \Eprint {https://arxiv.org/abs/0808.2096} {arXiv:0808.2096
  [hep-th]} \BibitemShut {NoStop}%
\bibitem [{\citenamefont {Shenker}\ and\ \citenamefont
  {Stanford}(2015)}]{Shenker:2014cwa}%
  \BibitemOpen
  \bibfield  {author} {\bibinfo {author} {\bibfnamefont {S.~H.}\ \bibnamefont
  {Shenker}}\ and\ \bibinfo {author} {\bibfnamefont {D.}~\bibnamefont
  {Stanford}},\ }\bibfield  {title} {\bibinfo {title} {{Stringy effects in
  scrambling}},\ }\href {https://doi.org/10.1007/JHEP05(2015)132} {\bibfield
  {journal} {\bibinfo  {journal} {JHEP}\ }\textbf {\bibinfo {volume} {05}},\
  \bibinfo {pages} {132}},\ \Eprint {https://arxiv.org/abs/1412.6087}
  {arXiv:1412.6087 [hep-th]} \BibitemShut {NoStop}%
\bibitem [{\citenamefont {Roberts}\ \emph {et~al.}(2015)\citenamefont
  {Roberts}, \citenamefont {Stanford},\ and\ \citenamefont
  {Susskind}}]{Roberts:2014isa}%
  \BibitemOpen
  \bibfield  {author} {\bibinfo {author} {\bibfnamefont {D.~A.}\ \bibnamefont
  {Roberts}}, \bibinfo {author} {\bibfnamefont {D.}~\bibnamefont {Stanford}},\
  and\ \bibinfo {author} {\bibfnamefont {L.}~\bibnamefont {Susskind}},\
  }\bibfield  {title} {\bibinfo {title} {{Localized shocks}},\ }\href
  {https://doi.org/10.1007/JHEP03(2015)051} {\bibfield  {journal} {\bibinfo
  {journal} {JHEP}\ }\textbf {\bibinfo {volume} {03}},\ \bibinfo {pages}
  {051}},\ \Eprint {https://arxiv.org/abs/1409.8180} {arXiv:1409.8180 [hep-th]}
  \BibitemShut {NoStop}%
\bibitem [{\citenamefont {Larkin}\ and\ \citenamefont
  {Ovchinnikov}(1969)}]{LaOv69}%
  \BibitemOpen
  \bibfield  {author} {\bibinfo {author} {\bibfnamefont {A.~I.}\ \bibnamefont
  {Larkin}}\ and\ \bibinfo {author} {\bibfnamefont {Y.~N.}\ \bibnamefont
  {Ovchinnikov}},\ }\bibfield  {title} {\bibinfo {title} {Quasiclassical method
  in the theory of superconductivity},\ }\href
  {http://www.jetp.ac.ru/cgi-bin/e/index/e/28/6/p1200?a=list} {\bibfield
  {journal} {\bibinfo  {journal} {Soviet Physics, JETP}\ }\textbf {\bibinfo
  {volume} {28}},\ \bibinfo {pages} {1200} (\bibinfo {year}
  {1969})}\BibitemShut {NoStop}%
\bibitem [{\citenamefont {Nahum}\ \emph {et~al.}(2018)\citenamefont {Nahum},
  \citenamefont {Vijay},\ and\ \citenamefont {Haah}}]{Nahum:2017yvy}%
  \BibitemOpen
  \bibfield  {author} {\bibinfo {author} {\bibfnamefont {A.}~\bibnamefont
  {Nahum}}, \bibinfo {author} {\bibfnamefont {S.}~\bibnamefont {Vijay}},\ and\
  \bibinfo {author} {\bibfnamefont {J.}~\bibnamefont {Haah}},\ }\bibfield
  {title} {\bibinfo {title} {{Operator Spreading in Random Unitary Circuits}},\
  }\href {https://doi.org/10.1103/PhysRevX.8.021014} {\bibfield  {journal}
  {\bibinfo  {journal} {Phys. Rev. X}\ }\textbf {\bibinfo {volume} {8}},\
  \bibinfo {pages} {021014} (\bibinfo {year} {2018})},\ \Eprint
  {https://arxiv.org/abs/1705.08975} {arXiv:1705.08975 [cond-mat.str-el]}
  \BibitemShut {NoStop}%
\bibitem [{\citenamefont {Qi}\ and\ \citenamefont
  {Streicher}(2019)}]{Qi:2018bje}%
  \BibitemOpen
  \bibfield  {author} {\bibinfo {author} {\bibfnamefont {X.-L.}\ \bibnamefont
  {Qi}}\ and\ \bibinfo {author} {\bibfnamefont {A.}~\bibnamefont {Streicher}},\
  }\bibfield  {title} {\bibinfo {title} {{Quantum Epidemiology: Operator
  Growth, Thermal Effects, and SYK}},\ }\href
  {https://doi.org/10.1007/JHEP08(2019)012} {\bibfield  {journal} {\bibinfo
  {journal} {JHEP}\ }\textbf {\bibinfo {volume} {08}},\ \bibinfo {pages}
  {012}},\ \Eprint {https://arxiv.org/abs/1810.11958} {arXiv:1810.11958
  [hep-th]} \BibitemShut {NoStop}%
\bibitem [{\citenamefont {Hunter-Jones}(2018)}]{Hunter-Jones:2018otn}%
  \BibitemOpen
  \bibfield  {author} {\bibinfo {author} {\bibfnamefont {N.}~\bibnamefont
  {Hunter-Jones}},\ }\bibfield  {title} {\bibinfo {title} {{Operator growth in
  random quantum circuits with symmetry}},\ }\href@noop {} {\  (\bibinfo {year}
  {2018})},\ \Eprint {https://arxiv.org/abs/1812.08219} {arXiv:1812.08219
  [quant-ph]} \BibitemShut {NoStop}%
\bibitem [{\citenamefont {von Keyserlingk}\ \emph {et~al.}(2018)\citenamefont
  {von Keyserlingk}, \citenamefont {Rakovszky}, \citenamefont {Pollmann},\ and\
  \citenamefont {Sondhi}}]{vonKeyserlingk:2017dyr}%
  \BibitemOpen
  \bibfield  {author} {\bibinfo {author} {\bibfnamefont {C.}~\bibnamefont {von
  Keyserlingk}}, \bibinfo {author} {\bibfnamefont {T.}~\bibnamefont
  {Rakovszky}}, \bibinfo {author} {\bibfnamefont {F.}~\bibnamefont
  {Pollmann}},\ and\ \bibinfo {author} {\bibfnamefont {S.}~\bibnamefont
  {Sondhi}},\ }\bibfield  {title} {\bibinfo {title} {{Operator hydrodynamics,
  OTOCs, and entanglement growth in systems without conservation laws}},\
  }\href {https://doi.org/10.1103/PhysRevX.8.021013} {\bibfield  {journal}
  {\bibinfo  {journal} {Phys. Rev. X}\ }\textbf {\bibinfo {volume} {8}},\
  \bibinfo {pages} {021013} (\bibinfo {year} {2018})},\ \Eprint
  {https://arxiv.org/abs/1705.08910} {arXiv:1705.08910 [cond-mat.str-el]}
  \BibitemShut {NoStop}%
\bibitem [{\citenamefont {Khemani}\ \emph {et~al.}(2018)\citenamefont
  {Khemani}, \citenamefont {Vishwanath},\ and\ \citenamefont
  {Huse}}]{PhysRevX.8.031057}%
  \BibitemOpen
  \bibfield  {author} {\bibinfo {author} {\bibfnamefont {V.}~\bibnamefont
  {Khemani}}, \bibinfo {author} {\bibfnamefont {A.}~\bibnamefont
  {Vishwanath}},\ and\ \bibinfo {author} {\bibfnamefont {D.~A.}\ \bibnamefont
  {Huse}},\ }\bibfield  {title} {\bibinfo {title} {Operator spreading and the
  emergence of dissipative hydrodynamics under unitary evolution with
  conservation laws},\ }\href {https://doi.org/10.1103/PhysRevX.8.031057}
  {\bibfield  {journal} {\bibinfo  {journal} {Phys. Rev. X}\ }\textbf {\bibinfo
  {volume} {8}},\ \bibinfo {pages} {031057} (\bibinfo {year}
  {2018})}\BibitemShut {NoStop}%
\bibitem [{\citenamefont {Dias}\ \emph {et~al.}(2021)\citenamefont {Dias},
  \citenamefont {Haque}, \citenamefont {Ribeiro},\ and\ \citenamefont
  {McClarty}}]{Dias:2021ncd}%
  \BibitemOpen
  \bibfield  {author} {\bibinfo {author} {\bibfnamefont {B.~C.}\ \bibnamefont
  {Dias}}, \bibinfo {author} {\bibfnamefont {M.}~\bibnamefont {Haque}},
  \bibinfo {author} {\bibfnamefont {P.}~\bibnamefont {Ribeiro}},\ and\ \bibinfo
  {author} {\bibfnamefont {P.}~\bibnamefont {McClarty}},\ }\bibfield  {title}
  {\bibinfo {title} {{Diffusive Operator Spreading for Random Unitary Free
  Fermion Circuits}},\ }\href@noop {} {\  (\bibinfo {year} {2021})},\ \Eprint
  {https://arxiv.org/abs/2102.09846} {arXiv:2102.09846 [cond-mat.str-el]}
  \BibitemShut {NoStop}%
\bibitem [{\citenamefont {Wu}\ \emph {et~al.}(2021)\citenamefont {Wu},
  \citenamefont {Zhang},\ and\ \citenamefont
  {Zhai}}]{PhysRevResearch.3.L032057}%
  \BibitemOpen
  \bibfield  {author} {\bibinfo {author} {\bibfnamefont {Y.}~\bibnamefont
  {Wu}}, \bibinfo {author} {\bibfnamefont {P.}~\bibnamefont {Zhang}},\ and\
  \bibinfo {author} {\bibfnamefont {H.}~\bibnamefont {Zhai}},\ }\bibfield
  {title} {\bibinfo {title} {Scrambling ability of quantum neural network
  architectures},\ }\href {https://doi.org/10.1103/PhysRevResearch.3.L032057}
  {\bibfield  {journal} {\bibinfo  {journal} {Phys. Rev. Research}\ }\textbf
  {\bibinfo {volume} {3}},\ \bibinfo {pages} {L032057} (\bibinfo {year}
  {2021})}\BibitemShut {NoStop}%
\bibitem [{\citenamefont {Roberts}\ \emph {et~al.}(2018)\citenamefont
  {Roberts}, \citenamefont {Stanford},\ and\ \citenamefont
  {Streicher}}]{Roberts:2018aa}%
  \BibitemOpen
  \bibfield  {author} {\bibinfo {author} {\bibfnamefont {D.~A.}\ \bibnamefont
  {Roberts}}, \bibinfo {author} {\bibfnamefont {D.}~\bibnamefont {Stanford}},\
  and\ \bibinfo {author} {\bibfnamefont {A.}~\bibnamefont {Streicher}},\
  }\bibfield  {title} {\bibinfo {title} {Operator growth in the syk model},\
  }\href {https://doi.org/10.1007/JHEP06(2018)122} {\bibfield  {journal}
  {\bibinfo  {journal} {Journal of High Energy Physics}\ }\textbf {\bibinfo
  {volume} {2018}},\ \bibinfo {pages} {122} (\bibinfo {year}
  {2018})}\BibitemShut {NoStop}%
\bibitem [{\citenamefont {Qi}\ \emph {et~al.}(2019)\citenamefont {Qi},
  \citenamefont {Davis}, \citenamefont {Periwal},\ and\ \citenamefont
  {Schleier-Smith}}]{qi2019}%
  \BibitemOpen
  \bibfield  {author} {\bibinfo {author} {\bibfnamefont {X.-L.}\ \bibnamefont
  {Qi}}, \bibinfo {author} {\bibfnamefont {E.~J.}\ \bibnamefont {Davis}},
  \bibinfo {author} {\bibfnamefont {A.}~\bibnamefont {Periwal}},\ and\ \bibinfo
  {author} {\bibfnamefont {M.}~\bibnamefont {Schleier-Smith}},\ }\href
  {https://arxiv.org/abs/1906.00524} {\bibinfo {title} {Measuring operator size
  growth in quantum quench experiments}} (\bibinfo {year} {2019}),\ \Eprint
  {https://arxiv.org/abs/1906.00524} {arXiv:1906.00524 [quant-ph]} \BibitemShut
  {NoStop}%
\bibitem [{\citenamefont {Lucas}\ and\ \citenamefont
  {Osborne}(2020)}]{Lucas:2020pgj}%
  \BibitemOpen
  \bibfield  {author} {\bibinfo {author} {\bibfnamefont {A.}~\bibnamefont
  {Lucas}}\ and\ \bibinfo {author} {\bibfnamefont {A.}~\bibnamefont
  {Osborne}},\ }\bibfield  {title} {\bibinfo {title} {{Operator growth bounds
  in a cartoon matrix model}},\ }\href {https://doi.org/10.1063/5.0022177}
  {\bibfield  {journal} {\bibinfo  {journal} {J. Math. Phys.}\ }\textbf
  {\bibinfo {volume} {61}},\ \bibinfo {pages} {122301} (\bibinfo {year}
  {2020})},\ \Eprint {https://arxiv.org/abs/2007.07165} {arXiv:2007.07165
  [hep-th]} \BibitemShut {NoStop}%
\bibitem [{\citenamefont {Lensky}\ \emph {et~al.}(2020)\citenamefont {Lensky},
  \citenamefont {Qi},\ and\ \citenamefont {Zhang}}]{Lensky:2020ubw}%
  \BibitemOpen
  \bibfield  {author} {\bibinfo {author} {\bibfnamefont {Y.~D.}\ \bibnamefont
  {Lensky}}, \bibinfo {author} {\bibfnamefont {X.-L.}\ \bibnamefont {Qi}},\
  and\ \bibinfo {author} {\bibfnamefont {P.}~\bibnamefont {Zhang}},\ }\bibfield
   {title} {\bibinfo {title} {{Size of bulk fermions in the SYK model}},\
  }\href {https://doi.org/10.1007/JHEP10(2020)053} {\bibfield  {journal}
  {\bibinfo  {journal} {JHEP}\ }\textbf {\bibinfo {volume} {10}},\ \bibinfo
  {pages} {053}},\ \Eprint {https://arxiv.org/abs/2002.01961} {arXiv:2002.01961
  [hep-th]} \BibitemShut {NoStop}%
\bibitem [{\citenamefont {Lucas}(2019)}]{PhysRevLett.122.216601}%
  \BibitemOpen
  \bibfield  {author} {\bibinfo {author} {\bibfnamefont {A.}~\bibnamefont
  {Lucas}},\ }\bibfield  {title} {\bibinfo {title} {Operator size at finite
  temperature and planckian bounds on quantum dynamics},\ }\href
  {https://doi.org/10.1103/PhysRevLett.122.216601} {\bibfield  {journal}
  {\bibinfo  {journal} {Phys. Rev. Lett.}\ }\textbf {\bibinfo {volume} {122}},\
  \bibinfo {pages} {216601} (\bibinfo {year} {2019})}\BibitemShut {NoStop}%
\bibitem [{\citenamefont {Chen}\ and\ \citenamefont
  {Lucas}(2021)}]{Chen:2019klo}%
  \BibitemOpen
  \bibfield  {author} {\bibinfo {author} {\bibfnamefont {C.-F.}\ \bibnamefont
  {Chen}}\ and\ \bibinfo {author} {\bibfnamefont {A.}~\bibnamefont {Lucas}},\
  }\bibfield  {title} {\bibinfo {title} {{Operator Growth Bounds from Graph
  Theory}},\ }\href {https://doi.org/10.1007/s00220-021-04151-6} {\bibfield
  {journal} {\bibinfo  {journal} {Commun. Math. Phys.}\ }\textbf {\bibinfo
  {volume} {385}},\ \bibinfo {pages} {1273} (\bibinfo {year} {2021})},\ \Eprint
  {https://arxiv.org/abs/1905.03682} {arXiv:1905.03682 [math-ph]} \BibitemShut
  {NoStop}%
\bibitem [{\citenamefont {Chen}\ \emph {et~al.}(2020)\citenamefont {Chen},
  \citenamefont {Gu},\ and\ \citenamefont {Lucas}}]{Chen:2020bmq}%
  \BibitemOpen
  \bibfield  {author} {\bibinfo {author} {\bibfnamefont {X.}~\bibnamefont
  {Chen}}, \bibinfo {author} {\bibfnamefont {Y.}~\bibnamefont {Gu}},\ and\
  \bibinfo {author} {\bibfnamefont {A.}~\bibnamefont {Lucas}},\ }\bibfield
  {title} {\bibinfo {title} {{Many-body quantum dynamics slows down at low
  density}},\ }\href {https://doi.org/10.21468/SciPostPhys.9.5.071} {\bibfield
  {journal} {\bibinfo  {journal} {SciPost Phys.}\ }\textbf {\bibinfo {volume}
  {9}},\ \bibinfo {pages} {071} (\bibinfo {year} {2020})},\ \Eprint
  {https://arxiv.org/abs/2007.10352} {arXiv:2007.10352 [quant-ph]} \BibitemShut
  {NoStop}%
\bibitem [{\citenamefont {Yin}\ and\ \citenamefont
  {Lucas}(2021)}]{Yin:2020oze}%
  \BibitemOpen
  \bibfield  {author} {\bibinfo {author} {\bibfnamefont {C.}~\bibnamefont
  {Yin}}\ and\ \bibinfo {author} {\bibfnamefont {A.}~\bibnamefont {Lucas}},\
  }\bibfield  {title} {\bibinfo {title} {{Quantum operator growth bounds for
  kicked tops and semiclassical spin chains}},\ }\href
  {https://doi.org/10.1103/PhysRevA.103.042414} {\bibfield  {journal} {\bibinfo
   {journal} {Phys. Rev. A}\ }\textbf {\bibinfo {volume} {103}},\ \bibinfo
  {pages} {042414} (\bibinfo {year} {2021})},\ \Eprint
  {https://arxiv.org/abs/2010.06592} {arXiv:2010.06592 [cond-mat.str-el]}
  \BibitemShut {NoStop}%
\bibitem [{\citenamefont {Zhou}\ and\ \citenamefont
  {Swingle}(2021)}]{Zhou:2021syv}%
  \BibitemOpen
  \bibfield  {author} {\bibinfo {author} {\bibfnamefont {T.}~\bibnamefont
  {Zhou}}\ and\ \bibinfo {author} {\bibfnamefont {B.}~\bibnamefont {Swingle}},\
  }\bibfield  {title} {\bibinfo {title} {{Operator Growth from Global
  Out-of-time-order Correlators}},\ }\href@noop {} {\  (\bibinfo {year}
  {2021})},\ \Eprint {https://arxiv.org/abs/2112.01562} {arXiv:2112.01562
  [quant-ph]} \BibitemShut {NoStop}%
\bibitem [{\citenamefont {Omanakuttan}\ \emph {et~al.}(2022)\citenamefont
  {Omanakuttan}, \citenamefont {Chinni}, \citenamefont {Blocher},\ and\
  \citenamefont {Poggi}}]{Omanakuttan:2022ikz}%
  \BibitemOpen
  \bibfield  {author} {\bibinfo {author} {\bibfnamefont {S.}~\bibnamefont
  {Omanakuttan}}, \bibinfo {author} {\bibfnamefont {K.}~\bibnamefont {Chinni}},
  \bibinfo {author} {\bibfnamefont {P.~D.}\ \bibnamefont {Blocher}},\ and\
  \bibinfo {author} {\bibfnamefont {P.~M.}\ \bibnamefont {Poggi}},\ }\bibfield
  {title} {\bibinfo {title} {{Scrambling and quantum chaos indicators from
  long-time properties of operator distributions}},\ }\href@noop {} {\
  (\bibinfo {year} {2022})},\ \Eprint {https://arxiv.org/abs/2211.15872}
  {arXiv:2211.15872 [quant-ph]} \BibitemShut {NoStop}%
\bibitem [{\citenamefont {Ippoliti}\ \emph {et~al.}(2023)\citenamefont
  {Ippoliti}, \citenamefont {Li}, \citenamefont {Rakovszky},\ and\
  \citenamefont {Khemani}}]{Ippoliti:2022vfn}%
  \BibitemOpen
  \bibfield  {author} {\bibinfo {author} {\bibfnamefont {M.}~\bibnamefont
  {Ippoliti}}, \bibinfo {author} {\bibfnamefont {Y.}~\bibnamefont {Li}},
  \bibinfo {author} {\bibfnamefont {T.}~\bibnamefont {Rakovszky}},\ and\
  \bibinfo {author} {\bibfnamefont {V.}~\bibnamefont {Khemani}},\ }\bibfield
  {title} {\bibinfo {title} {{Operator Relaxation and the Optimal Depth of
  Classical Shadows}},\ }\href {https://doi.org/10.1103/PhysRevLett.130.230403}
  {\bibfield  {journal} {\bibinfo  {journal} {Phys. Rev. Lett.}\ }\textbf
  {\bibinfo {volume} {130}},\ \bibinfo {pages} {230403} (\bibinfo {year}
  {2023})},\ \Eprint {https://arxiv.org/abs/2212.11963} {arXiv:2212.11963
  [quant-ph]} \BibitemShut {NoStop}%
\bibitem [{\citenamefont {Xu}\ and\ \citenamefont
  {Swingle}(2022)}]{Xu:2022vko}%
  \BibitemOpen
  \bibfield  {author} {\bibinfo {author} {\bibfnamefont {S.}~\bibnamefont
  {Xu}}\ and\ \bibinfo {author} {\bibfnamefont {B.}~\bibnamefont {Swingle}},\
  }\bibfield  {title} {\bibinfo {title} {{Scrambling Dynamics and Out-of-Time
  Ordered Correlators in Quantum Many-Body Systems: a Tutorial}},\ }\href@noop
  {} {\  (\bibinfo {year} {2022})},\ \Eprint {https://arxiv.org/abs/2202.07060}
  {arXiv:2202.07060 [quant-ph]} \BibitemShut {NoStop}%
\bibitem [{\citenamefont {Bhattacharyya}\ \emph {et~al.}(2022)\citenamefont
  {Bhattacharyya}, \citenamefont {Joshi},\ and\ \citenamefont
  {Sundar}}]{Bhattacharyya:2021ypq}%
  \BibitemOpen
  \bibfield  {author} {\bibinfo {author} {\bibfnamefont {A.}~\bibnamefont
  {Bhattacharyya}}, \bibinfo {author} {\bibfnamefont {L.~K.}\ \bibnamefont
  {Joshi}},\ and\ \bibinfo {author} {\bibfnamefont {B.}~\bibnamefont
  {Sundar}},\ }\bibfield  {title} {\bibinfo {title} {{Quantum information
  scrambling: from holography to quantum simulators}},\ }\href
  {https://doi.org/10.1140/epjc/s10052-022-10377-y} {\bibfield  {journal}
  {\bibinfo  {journal} {Eur. Phys. J. C}\ }\textbf {\bibinfo {volume} {82}},\
  \bibinfo {pages} {458} (\bibinfo {year} {2022})},\ \Eprint
  {https://arxiv.org/abs/2111.11945} {arXiv:2111.11945 [hep-th]} \BibitemShut
  {NoStop}%
\bibitem [{\citenamefont {Hosur}\ \emph {et~al.}(2016)\citenamefont {Hosur},
  \citenamefont {Qi}, \citenamefont {Roberts},\ and\ \citenamefont
  {Yoshida}}]{Hosur:2015ylk}%
  \BibitemOpen
  \bibfield  {author} {\bibinfo {author} {\bibfnamefont {P.}~\bibnamefont
  {Hosur}}, \bibinfo {author} {\bibfnamefont {X.-L.}\ \bibnamefont {Qi}},
  \bibinfo {author} {\bibfnamefont {D.~A.}\ \bibnamefont {Roberts}},\ and\
  \bibinfo {author} {\bibfnamefont {B.}~\bibnamefont {Yoshida}},\ }\bibfield
  {title} {\bibinfo {title} {{Chaos in quantum channels}},\ }\href
  {https://doi.org/10.1007/JHEP02(2016)004} {\bibfield  {journal} {\bibinfo
  {journal} {JHEP}\ }\textbf {\bibinfo {volume} {02}},\ \bibinfo {pages}
  {004}},\ \Eprint {https://arxiv.org/abs/1511.04021} {arXiv:1511.04021
  [hep-th]} \BibitemShut {NoStop}%
\bibitem [{\citenamefont {Fan}\ \emph {et~al.}(2017)\citenamefont {Fan},
  \citenamefont {Zhang}, \citenamefont {Shen},\ and\ \citenamefont
  {Zhai}}]{Fan:2016ean}%
  \BibitemOpen
  \bibfield  {author} {\bibinfo {author} {\bibfnamefont {R.}~\bibnamefont
  {Fan}}, \bibinfo {author} {\bibfnamefont {P.}~\bibnamefont {Zhang}}, \bibinfo
  {author} {\bibfnamefont {H.}~\bibnamefont {Shen}},\ and\ \bibinfo {author}
  {\bibfnamefont {H.}~\bibnamefont {Zhai}},\ }\bibfield  {title} {\bibinfo
  {title} {{Out-of-Time-Order Correlation for Many-Body Localization}},\ }\href
  {https://doi.org/10.1016/j.scib.2017.04.011} {\bibfield  {journal} {\bibinfo
  {journal} {Sci. Bull.}\ }\textbf {\bibinfo {volume} {62}},\ \bibinfo {pages}
  {707} (\bibinfo {year} {2017})},\ \Eprint {https://arxiv.org/abs/1608.01914}
  {arXiv:1608.01914 [cond-mat.quant-gas]} \BibitemShut {NoStop}%
\bibitem [{\citenamefont {Padmanabhan}\ \emph {et~al.}(2017)\citenamefont
  {Padmanabhan}, \citenamefont {Rey}, \citenamefont {Teixeira},\ and\
  \citenamefont {Trancanelli}}]{Padmanabhan:2017ekk}%
  \BibitemOpen
  \bibfield  {author} {\bibinfo {author} {\bibfnamefont {P.}~\bibnamefont
  {Padmanabhan}}, \bibinfo {author} {\bibfnamefont {S.-J.}\ \bibnamefont
  {Rey}}, \bibinfo {author} {\bibfnamefont {D.}~\bibnamefont {Teixeira}},\ and\
  \bibinfo {author} {\bibfnamefont {D.}~\bibnamefont {Trancanelli}},\
  }\bibfield  {title} {\bibinfo {title} {{Supersymmetric many-body systems from
  partial symmetries \textemdash{} integrability, localization and
  scrambling}},\ }\href {https://doi.org/10.1007/JHEP05(2017)136} {\bibfield
  {journal} {\bibinfo  {journal} {JHEP}\ }\textbf {\bibinfo {volume} {05}},\
  \bibinfo {pages} {136}},\ \Eprint {https://arxiv.org/abs/1702.02091}
  {arXiv:1702.02091 [hep-th]} \BibitemShut {NoStop}%
\bibitem [{\citenamefont {Bergamasco}\ \emph {et~al.}(2019)\citenamefont
  {Bergamasco}, \citenamefont {Carlo},\ and\ \citenamefont
  {Rivas}}]{PhysRevResearch.1.033044}%
  \BibitemOpen
  \bibfield  {author} {\bibinfo {author} {\bibfnamefont {P.~D.}\ \bibnamefont
  {Bergamasco}}, \bibinfo {author} {\bibfnamefont {G.~G.}\ \bibnamefont
  {Carlo}},\ and\ \bibinfo {author} {\bibfnamefont {A.~M.~F.}\ \bibnamefont
  {Rivas}},\ }\bibfield  {title} {\bibinfo {title} {Out-of-time ordered
  correlators, complexity, and entropy in bipartite systems},\ }\href
  {https://doi.org/10.1103/PhysRevResearch.1.033044} {\bibfield  {journal}
  {\bibinfo  {journal} {Phys. Rev. Res.}\ }\textbf {\bibinfo {volume} {1}},\
  \bibinfo {pages} {033044} (\bibinfo {year} {2019})}\BibitemShut {NoStop}%
\bibitem [{\citenamefont {Bergamasco}\ \emph {et~al.}(2020)\citenamefont
  {Bergamasco}, \citenamefont {Carlo},\ and\ \citenamefont
  {Rivas}}]{PhysRevE.102.052133}%
  \BibitemOpen
  \bibfield  {author} {\bibinfo {author} {\bibfnamefont {P.~D.}\ \bibnamefont
  {Bergamasco}}, \bibinfo {author} {\bibfnamefont {G.~G.}\ \bibnamefont
  {Carlo}},\ and\ \bibinfo {author} {\bibfnamefont {A.~M.~F.}\ \bibnamefont
  {Rivas}},\ }\bibfield  {title} {\bibinfo {title} {Relevant out-of-time-order
  correlator operators: Footprints of the classical dynamics},\ }\href
  {https://doi.org/10.1103/PhysRevE.102.052133} {\bibfield  {journal} {\bibinfo
   {journal} {Phys. Rev. E}\ }\textbf {\bibinfo {volume} {102}},\ \bibinfo
  {pages} {052133} (\bibinfo {year} {2020})}\BibitemShut {NoStop}%
\bibitem [{\citenamefont {Chen}(2021)}]{Chen:2020atj}%
  \BibitemOpen
  \bibfield  {author} {\bibinfo {author} {\bibfnamefont {Y.}~\bibnamefont
  {Chen}},\ }\bibfield  {title} {\bibinfo {title} {{Entropy linear response
  theory with non-Markovian bath}},\ }\href
  {https://doi.org/10.1007/JHEP04(2021)215} {\bibfield  {journal} {\bibinfo
  {journal} {JHEP}\ }\textbf {\bibinfo {volume} {04}},\ \bibinfo {pages}
  {215}},\ \Eprint {https://arxiv.org/abs/2012.00223} {arXiv:2012.00223
  [hep-th]} \BibitemShut {NoStop}%
\bibitem [{\citenamefont {Li}\ \emph {et~al.}(2017)\citenamefont {Li},
  \citenamefont {Fan}, \citenamefont {Wang}, \citenamefont {Ye}, \citenamefont
  {Zeng}, \citenamefont {Zhai}, \citenamefont {Peng},\ and\ \citenamefont
  {Du}}]{PhysRevX.7.031011}%
  \BibitemOpen
  \bibfield  {author} {\bibinfo {author} {\bibfnamefont {J.}~\bibnamefont
  {Li}}, \bibinfo {author} {\bibfnamefont {R.}~\bibnamefont {Fan}}, \bibinfo
  {author} {\bibfnamefont {H.}~\bibnamefont {Wang}}, \bibinfo {author}
  {\bibfnamefont {B.}~\bibnamefont {Ye}}, \bibinfo {author} {\bibfnamefont
  {B.}~\bibnamefont {Zeng}}, \bibinfo {author} {\bibfnamefont {H.}~\bibnamefont
  {Zhai}}, \bibinfo {author} {\bibfnamefont {X.}~\bibnamefont {Peng}},\ and\
  \bibinfo {author} {\bibfnamefont {J.}~\bibnamefont {Du}},\ }\bibfield
  {title} {\bibinfo {title} {Measuring out-of-time-order correlators on a
  nuclear magnetic resonance quantum simulator},\ }\href
  {https://doi.org/10.1103/PhysRevX.7.031011} {\bibfield  {journal} {\bibinfo
  {journal} {Phys. Rev. X}\ }\textbf {\bibinfo {volume} {7}},\ \bibinfo {pages}
  {031011} (\bibinfo {year} {2017})}\BibitemShut {NoStop}%
\bibitem [{\citenamefont {Dadras}\ and\ \citenamefont
  {Kitaev}(2021)}]{dadras2021perturbative}%
  \BibitemOpen
  \bibfield  {author} {\bibinfo {author} {\bibfnamefont {P.}~\bibnamefont
  {Dadras}}\ and\ \bibinfo {author} {\bibfnamefont {A.}~\bibnamefont
  {Kitaev}},\ }\bibfield  {title} {\bibinfo {title} {{Perturbative calculations
  of entanglement entropy}},\ }\href {https://doi.org/10.1007/JHEP03(2021)198}
  {\bibfield  {journal} {\bibinfo  {journal} {JHEP}\ }\textbf {\bibinfo
  {volume} {03}},\ \bibinfo {pages} {198}},\ \Eprint
  {https://arxiv.org/abs/2011.09622} {arXiv:2011.09622 [hep-th]} \BibitemShut
  {NoStop}%
\bibitem [{\citenamefont {Zhang}(2023)}]{Zhang:2023gzl}%
  \BibitemOpen
  \bibfield  {author} {\bibinfo {author} {\bibfnamefont {P.}~\bibnamefont
  {Zhang}},\ }\bibfield  {title} {\bibinfo {title} {{Perturbative Page curve
  induced by external impulse}},\ }\href
  {https://doi.org/10.1007/JHEP09(2023)056} {\bibfield  {journal} {\bibinfo
  {journal} {JHEP}\ }\textbf {\bibinfo {volume} {09}},\ \bibinfo {pages}
  {056}},\ \Eprint {https://arxiv.org/abs/2305.18329} {arXiv:2305.18329
  [cond-mat.stat-mech]} \BibitemShut {NoStop}%
\bibitem [{\citenamefont {Chen}\ \emph {et~al.}(2017)\citenamefont {Chen},
  \citenamefont {Zhai},\ and\ \citenamefont {Zhang}}]{Chen:2017dbb}%
  \BibitemOpen
  \bibfield  {author} {\bibinfo {author} {\bibfnamefont {Y.}~\bibnamefont
  {Chen}}, \bibinfo {author} {\bibfnamefont {H.}~\bibnamefont {Zhai}},\ and\
  \bibinfo {author} {\bibfnamefont {P.}~\bibnamefont {Zhang}},\ }\bibfield
  {title} {\bibinfo {title} {{Tunable Quantum Chaos in the Sachdev-Ye-Kitaev
  Model Coupled to a Thermal Bath}},\ }\href
  {https://doi.org/10.1007/JHEP07(2017)150} {\bibfield  {journal} {\bibinfo
  {journal} {JHEP}\ }\textbf {\bibinfo {volume} {07}},\ \bibinfo {pages}
  {150}},\ \Eprint {https://arxiv.org/abs/1705.09818} {arXiv:1705.09818
  [hep-th]} \BibitemShut {NoStop}%
\bibitem [{\citenamefont {Zhang}(2019)}]{Zhang:2019fcy}%
  \BibitemOpen
  \bibfield  {author} {\bibinfo {author} {\bibfnamefont {P.}~\bibnamefont
  {Zhang}},\ }\bibfield  {title} {\bibinfo {title} {{Evaporation dynamics of
  the Sachdev-Ye-Kitaev model}},\ }\href
  {https://doi.org/10.1103/PhysRevB.100.245104} {\bibfield  {journal} {\bibinfo
   {journal} {Phys. Rev. B}\ }\textbf {\bibinfo {volume} {100}},\ \bibinfo
  {pages} {245104} (\bibinfo {year} {2019})},\ \Eprint
  {https://arxiv.org/abs/1909.10637} {arXiv:1909.10637 [cond-mat.str-el]}
  \BibitemShut {NoStop}%
\bibitem [{\citenamefont {Syzranov}\ \emph {et~al.}(2018)\citenamefont
  {Syzranov}, \citenamefont {Gorshkov},\ and\ \citenamefont
  {Galitski}}]{PhysRevB.97.161114}%
  \BibitemOpen
  \bibfield  {author} {\bibinfo {author} {\bibfnamefont {S.~V.}\ \bibnamefont
  {Syzranov}}, \bibinfo {author} {\bibfnamefont {A.~V.}\ \bibnamefont
  {Gorshkov}},\ and\ \bibinfo {author} {\bibfnamefont {V.}~\bibnamefont
  {Galitski}},\ }\bibfield  {title} {\bibinfo {title} {Out-of-time-order
  correlators in finite open systems},\ }\href
  {https://doi.org/10.1103/PhysRevB.97.161114} {\bibfield  {journal} {\bibinfo
  {journal} {Phys. Rev. B}\ }\textbf {\bibinfo {volume} {97}},\ \bibinfo
  {pages} {161114} (\bibinfo {year} {2018})}\BibitemShut {NoStop}%
\bibitem [{\citenamefont {Tuziemski}(2019)}]{PhysRevA.100.062106}%
  \BibitemOpen
  \bibfield  {author} {\bibinfo {author} {\bibfnamefont {J.}~\bibnamefont
  {Tuziemski}},\ }\bibfield  {title} {\bibinfo {title} {Out-of-time-ordered
  correlation functions in open systems: A feynman-vernon influence functional
  approach},\ }\href {https://doi.org/10.1103/PhysRevA.100.062106} {\bibfield
  {journal} {\bibinfo  {journal} {Phys. Rev. A}\ }\textbf {\bibinfo {volume}
  {100}},\ \bibinfo {pages} {062106} (\bibinfo {year} {2019})}\BibitemShut
  {NoStop}%
\bibitem [{\citenamefont {Almheiri}\ \emph {et~al.}(2019)\citenamefont
  {Almheiri}, \citenamefont {Milekhin},\ and\ \citenamefont
  {Swingle}}]{Almheiri:2019jqq}%
  \BibitemOpen
  \bibfield  {author} {\bibinfo {author} {\bibfnamefont {A.}~\bibnamefont
  {Almheiri}}, \bibinfo {author} {\bibfnamefont {A.}~\bibnamefont {Milekhin}},\
  and\ \bibinfo {author} {\bibfnamefont {B.}~\bibnamefont {Swingle}},\
  }\bibfield  {title} {\bibinfo {title} {{Universal Constraints on Energy Flow
  and SYK Thermalization}},\ }\href@noop {} {\  (\bibinfo {year} {2019})},\
  \Eprint {https://arxiv.org/abs/1912.04912} {arXiv:1912.04912 [hep-th]}
  \BibitemShut {NoStop}%
\bibitem [{\citenamefont {Zhang}\ and\ \citenamefont
  {Yu}(2023)}]{Zhang:2023xrr}%
  \BibitemOpen
  \bibfield  {author} {\bibinfo {author} {\bibfnamefont {P.}~\bibnamefont
  {Zhang}}\ and\ \bibinfo {author} {\bibfnamefont {Z.}~\bibnamefont {Yu}},\
  }\bibfield  {title} {\bibinfo {title} {{Dynamical Transition of Operator Size
  Growth in Quantum Systems Embedded in an Environment}},\ }\href
  {https://doi.org/10.1103/PhysRevLett.130.250401} {\bibfield  {journal}
  {\bibinfo  {journal} {Phys. Rev. Lett.}\ }\textbf {\bibinfo {volume} {130}},\
  \bibinfo {pages} {250401} (\bibinfo {year} {2023})}\BibitemShut {NoStop}%
\bibitem [{\citenamefont {Weinstein}\ \emph {et~al.}(2023)\citenamefont
  {Weinstein}, \citenamefont {Kelly}, \citenamefont {Marino},\ and\
  \citenamefont {Altman}}]{Weinstein:2022yce}%
  \BibitemOpen
  \bibfield  {author} {\bibinfo {author} {\bibfnamefont {Z.}~\bibnamefont
  {Weinstein}}, \bibinfo {author} {\bibfnamefont {S.~P.}\ \bibnamefont
  {Kelly}}, \bibinfo {author} {\bibfnamefont {J.}~\bibnamefont {Marino}},\ and\
  \bibinfo {author} {\bibfnamefont {E.}~\bibnamefont {Altman}},\ }\bibfield
  {title} {\bibinfo {title} {{Scrambling Transition in a Radiative Random
  Unitary Circuit}},\ }\href {https://doi.org/10.1103/PhysRevLett.131.220404}
  {\bibfield  {journal} {\bibinfo  {journal} {Phys. Rev. Lett.}\ }\textbf
  {\bibinfo {volume} {131}},\ \bibinfo {pages} {220404} (\bibinfo {year}
  {2023})},\ \Eprint {https://arxiv.org/abs/2210.14242} {arXiv:2210.14242
  [quant-ph]} \BibitemShut {NoStop}%
\bibitem [{\citenamefont {Liu}\ \emph {et~al.}(2023)\citenamefont {Liu},
  \citenamefont {Tang},\ and\ \citenamefont {Zhai}}]{PhysRevResearch.5.033085}%
  \BibitemOpen
  \bibfield  {author} {\bibinfo {author} {\bibfnamefont {C.}~\bibnamefont
  {Liu}}, \bibinfo {author} {\bibfnamefont {H.}~\bibnamefont {Tang}},\ and\
  \bibinfo {author} {\bibfnamefont {H.}~\bibnamefont {Zhai}},\ }\bibfield
  {title} {\bibinfo {title} {Krylov complexity in open quantum systems},\
  }\href {https://doi.org/10.1103/PhysRevResearch.5.033085} {\bibfield
  {journal} {\bibinfo  {journal} {Phys. Rev. Res.}\ }\textbf {\bibinfo {volume}
  {5}},\ \bibinfo {pages} {033085} (\bibinfo {year} {2023})}\BibitemShut
  {NoStop}%
\bibitem [{\citenamefont {Bhattacharya}\ \emph {et~al.}(2022)\citenamefont
  {Bhattacharya}, \citenamefont {Nandy}, \citenamefont {Nath},\ and\
  \citenamefont {Sahu}}]{Bhattacharya:2022gbz}%
  \BibitemOpen
  \bibfield  {author} {\bibinfo {author} {\bibfnamefont {A.}~\bibnamefont
  {Bhattacharya}}, \bibinfo {author} {\bibfnamefont {P.}~\bibnamefont {Nandy}},
  \bibinfo {author} {\bibfnamefont {P.~P.}\ \bibnamefont {Nath}},\ and\
  \bibinfo {author} {\bibfnamefont {H.}~\bibnamefont {Sahu}},\ }\bibfield
  {title} {\bibinfo {title} {{Operator growth and Krylov construction in
  dissipative open quantum systems}},\ }\href
  {https://doi.org/10.1007/JHEP12(2022)081} {\bibfield  {journal} {\bibinfo
  {journal} {JHEP}\ }\textbf {\bibinfo {volume} {12}},\ \bibinfo {pages}
  {081}},\ \Eprint {https://arxiv.org/abs/2207.05347} {arXiv:2207.05347
  [quant-ph]} \BibitemShut {NoStop}%
\bibitem [{\citenamefont {Schuster}\ and\ \citenamefont
  {Yao}(2023)}]{Schuster:2022bot}%
  \BibitemOpen
  \bibfield  {author} {\bibinfo {author} {\bibfnamefont {T.}~\bibnamefont
  {Schuster}}\ and\ \bibinfo {author} {\bibfnamefont {N.~Y.}\ \bibnamefont
  {Yao}},\ }\bibfield  {title} {\bibinfo {title} {{Operator Growth in Open
  Quantum Systems}},\ }\href {https://doi.org/10.1103/PhysRevLett.131.160402}
  {\bibfield  {journal} {\bibinfo  {journal} {Phys. Rev. Lett.}\ }\textbf
  {\bibinfo {volume} {131}},\ \bibinfo {pages} {160402} (\bibinfo {year}
  {2023})},\ \Eprint {https://arxiv.org/abs/2208.12272} {arXiv:2208.12272
  [quant-ph]} \BibitemShut {NoStop}%
\bibitem [{\citenamefont {Bhattacharjee}\ \emph {et~al.}(2023)\citenamefont
  {Bhattacharjee}, \citenamefont {Cao}, \citenamefont {Nandy},\ and\
  \citenamefont {Pathak}}]{Bhattacharjee:2022lzy}%
  \BibitemOpen
  \bibfield  {author} {\bibinfo {author} {\bibfnamefont {B.}~\bibnamefont
  {Bhattacharjee}}, \bibinfo {author} {\bibfnamefont {X.}~\bibnamefont {Cao}},
  \bibinfo {author} {\bibfnamefont {P.}~\bibnamefont {Nandy}},\ and\ \bibinfo
  {author} {\bibfnamefont {T.}~\bibnamefont {Pathak}},\ }\bibfield  {title}
  {\bibinfo {title} {{Operator growth in open quantum systems: lessons from the
  dissipative SYK}},\ }\href {https://doi.org/10.1007/JHEP03(2023)054}
  {\bibfield  {journal} {\bibinfo  {journal} {JHEP}\ }\textbf {\bibinfo
  {volume} {03}},\ \bibinfo {pages} {054}},\ \Eprint
  {https://arxiv.org/abs/2212.06180} {arXiv:2212.06180 [quant-ph]} \BibitemShut
  {NoStop}%
\bibitem [{\citenamefont {Bhattacharjee}\ \emph {et~al.}(2024)\citenamefont
  {Bhattacharjee}, \citenamefont {Nandy},\ and\ \citenamefont
  {Pathak}}]{Bhattacharjee:2023uwx}%
  \BibitemOpen
  \bibfield  {author} {\bibinfo {author} {\bibfnamefont {B.}~\bibnamefont
  {Bhattacharjee}}, \bibinfo {author} {\bibfnamefont {P.}~\bibnamefont
  {Nandy}},\ and\ \bibinfo {author} {\bibfnamefont {T.}~\bibnamefont
  {Pathak}},\ }\bibfield  {title} {\bibinfo {title} {{Operator dynamics in
  Lindbladian SYK: a Krylov complexity perspective}},\ }\href
  {https://doi.org/10.1007/JHEP01(2024)094} {\bibfield  {journal} {\bibinfo
  {journal} {JHEP}\ }\textbf {\bibinfo {volume} {01}},\ \bibinfo {pages}
  {094}},\ \Eprint {https://arxiv.org/abs/2311.00753} {arXiv:2311.00753
  [quant-ph]} \BibitemShut {NoStop}%
\bibitem [{\citenamefont {Garc\'{\i}a-Garc\'{\i}a}\ \emph
  {et~al.}(2024)\citenamefont {Garc\'{\i}a-Garc\'{\i}a}, \citenamefont
  {Verbaarschot},\ and\ \citenamefont {Zheng}}]{PhysRevD.110.086010}%
  \BibitemOpen
  \bibfield  {author} {\bibinfo {author} {\bibfnamefont {A.~M.}\ \bibnamefont
  {Garc\'{\i}a-Garc\'{\i}a}}, \bibinfo {author} {\bibfnamefont {J.~J.~M.}\
  \bibnamefont {Verbaarschot}},\ and\ \bibinfo {author} {\bibfnamefont {J.-p.}\
  \bibnamefont {Zheng}},\ }\bibfield  {title} {\bibinfo {title} {Lyapunov
  exponent as a signature of dissipative many-body quantum chaos},\ }\href
  {https://doi.org/10.1103/PhysRevD.110.086010} {\bibfield  {journal} {\bibinfo
   {journal} {Phys. Rev. D}\ }\textbf {\bibinfo {volume} {110}},\ \bibinfo
  {pages} {086010} (\bibinfo {year} {2024})}\BibitemShut {NoStop}%
\bibitem [{\citenamefont {Zhang}\ and\ \citenamefont
  {Yu}(2024)}]{Zhang:2024vsa}%
  \BibitemOpen
  \bibfield  {author} {\bibinfo {author} {\bibfnamefont {P.}~\bibnamefont
  {Zhang}}\ and\ \bibinfo {author} {\bibfnamefont {Z.}~\bibnamefont {Yu}},\
  }\bibfield  {title} {\bibinfo {title} {{Environment-induced information
  scrambling transition with charge conservations}},\ }\href
  {https://doi.org/10.1007/s43673-024-00124-8} {\bibfield  {journal} {\bibinfo
  {journal} {AAPPS Bull.}\ }\textbf {\bibinfo {volume} {34}},\ \bibinfo {pages}
  {19} (\bibinfo {year} {2024})},\ \Eprint {https://arxiv.org/abs/2403.08622}
  {arXiv:2403.08622 [quant-ph]} \BibitemShut {NoStop}%
\bibitem [{\citenamefont {Zhou}\ \emph {et~al.}(2024)\citenamefont {Zhou},
  \citenamefont {Zhang},\ and\ \citenamefont {Yu}}]{Zhou:2024osg}%
  \BibitemOpen
  \bibfield  {author} {\bibinfo {author} {\bibfnamefont {S.}~\bibnamefont
  {Zhou}}, \bibinfo {author} {\bibfnamefont {P.}~\bibnamefont {Zhang}},\ and\
  \bibinfo {author} {\bibfnamefont {Z.}~\bibnamefont {Yu}},\ }\bibfield
  {title} {\bibinfo {title} {{Environment-induced Transitions in Many-body
  Quantum Teleportation}},\ }\href@noop {} {\  (\bibinfo {year} {2024})},\
  \Eprint {https://arxiv.org/abs/2406.02277} {arXiv:2406.02277 [quant-ph]}
  \BibitemShut {NoStop}%
\bibitem [{\citenamefont {Aleiner}\ \emph {et~al.}(2016)\citenamefont
  {Aleiner}, \citenamefont {Faoro},\ and\ \citenamefont
  {Ioffe}}]{aleiner2016microscopic}%
  \BibitemOpen
  \bibfield  {author} {\bibinfo {author} {\bibfnamefont {I.~L.}\ \bibnamefont
  {Aleiner}}, \bibinfo {author} {\bibfnamefont {L.}~\bibnamefont {Faoro}},\
  and\ \bibinfo {author} {\bibfnamefont {L.~B.}\ \bibnamefont {Ioffe}},\
  }\bibfield  {title} {\bibinfo {title} {Microscopic model of quantum butterfly
  effect: out-of-time-order correlators and traveling combustion waves},\
  }\href@noop {} {\bibfield  {journal} {\bibinfo  {journal} {Annals of
  Physics}\ }\textbf {\bibinfo {volume} {375}},\ \bibinfo {pages} {378}
  (\bibinfo {year} {2016})}\BibitemShut {NoStop}%
\bibitem [{Note1()}]{Note1}%
  \BibitemOpen
  \bibinfo {note} {Considering an initial state without system-probe
  interaction leads only to a result differing by $O(\kappa ^2)$.}\BibitemShut
  {Stop}%
\bibitem [{\citenamefont {Page}(1993)}]{Page:1993df}%
  \BibitemOpen
  \bibfield  {author} {\bibinfo {author} {\bibfnamefont {D.~N.}\ \bibnamefont
  {Page}},\ }\bibfield  {title} {\bibinfo {title} {{Average entropy of a
  subsystem}},\ }\href {https://doi.org/10.1103/PhysRevLett.71.1291} {\bibfield
   {journal} {\bibinfo  {journal} {Phys. Rev. Lett.}\ }\textbf {\bibinfo
  {volume} {71}},\ \bibinfo {pages} {1291} (\bibinfo {year} {1993})},\ \Eprint
  {https://arxiv.org/abs/gr-qc/9305007} {arXiv:gr-qc/9305007} \BibitemShut
  {NoStop}%
\bibitem [{\citenamefont {Maldacena}\ \emph {et~al.}(2016)\citenamefont
  {Maldacena}, \citenamefont {Shenker},\ and\ \citenamefont
  {Stanford}}]{maldacena2016bound}%
  \BibitemOpen
  \bibfield  {author} {\bibinfo {author} {\bibfnamefont {J.}~\bibnamefont
  {Maldacena}}, \bibinfo {author} {\bibfnamefont {S.~H.}\ \bibnamefont
  {Shenker}},\ and\ \bibinfo {author} {\bibfnamefont {D.}~\bibnamefont
  {Stanford}},\ }\bibfield  {title} {\bibinfo {title} {{A bound on chaos}},\
  }\href {https://doi.org/10.1007/JHEP08(2016)106} {\bibfield  {journal}
  {\bibinfo  {journal} {JHEP}\ }\textbf {\bibinfo {volume} {08}},\ \bibinfo
  {pages} {106}},\ \Eprint {https://arxiv.org/abs/1503.01409} {arXiv:1503.01409
  [hep-th]} \BibitemShut {NoStop}%
\bibitem [{\citenamefont {Gu}\ \emph {et~al.}(2022)\citenamefont {Gu},
  \citenamefont {Kitaev},\ and\ \citenamefont {Zhang}}]{gu2022two}%
  \BibitemOpen
  \bibfield  {author} {\bibinfo {author} {\bibfnamefont {Y.}~\bibnamefont
  {Gu}}, \bibinfo {author} {\bibfnamefont {A.}~\bibnamefont {Kitaev}},\ and\
  \bibinfo {author} {\bibfnamefont {P.}~\bibnamefont {Zhang}},\ }\bibfield
  {title} {\bibinfo {title} {{A two-way approach to out-of-time-order
  correlators}},\ }\href {https://doi.org/10.1007/JHEP03(2022)133} {\bibfield
  {journal} {\bibinfo  {journal} {JHEP}\ }\textbf {\bibinfo {volume} {03}},\
  \bibinfo {pages} {133}},\ \Eprint {https://arxiv.org/abs/2111.12007}
  {arXiv:2111.12007 [hep-th]} \BibitemShut {NoStop}%
\bibitem [{\citenamefont {Stanford}\ \emph {et~al.}(2022)\citenamefont
  {Stanford}, \citenamefont {Yang},\ and\ \citenamefont
  {Yao}}]{Stanford:2021bhl}%
  \BibitemOpen
  \bibfield  {author} {\bibinfo {author} {\bibfnamefont {D.}~\bibnamefont
  {Stanford}}, \bibinfo {author} {\bibfnamefont {Z.}~\bibnamefont {Yang}},\
  and\ \bibinfo {author} {\bibfnamefont {S.}~\bibnamefont {Yao}},\ }\bibfield
  {title} {\bibinfo {title} {{Subleading Weingartens}},\ }\href
  {https://doi.org/10.1007/JHEP02(2022)200} {\bibfield  {journal} {\bibinfo
  {journal} {JHEP}\ }\textbf {\bibinfo {volume} {02}},\ \bibinfo {pages}
  {200}},\ \Eprint {https://arxiv.org/abs/2107.10252} {arXiv:2107.10252
  [hep-th]} \BibitemShut {NoStop}%
\bibitem [{\citenamefont {Zhang}\ and\ \citenamefont
  {Gu}(2022)}]{sizenewpaper}%
  \BibitemOpen
  \bibfield  {author} {\bibinfo {author} {\bibfnamefont {P.}~\bibnamefont
  {Zhang}}\ and\ \bibinfo {author} {\bibfnamefont {Y.}~\bibnamefont {Gu}},\
  }\bibfield  {title} {\bibinfo {title} {{Operator Size Distribution in Large
  $N$ Quantum Mechanics of Majorana Fermions}},\ }\href@noop {} {\  (\bibinfo
  {year} {2022})},\ \Eprint {https://arxiv.org/abs/2212.04358}
  {arXiv:2212.04358 [cond-mat.str-el]} \BibitemShut {NoStop}%
\bibitem [{Note2()}]{Note2}%
  \BibitemOpen
  \bibinfo {note} {Equivalently, we can choose $X = i\eta (c_{S,-\protect
  \mathbf {k}}^1+c_{S,\protect \mathbf {k}}^{1,\dagger })$, with the auxiliary
  Majorana fermion mode $\eta $, to make the perturbation operator $X$
  bosonic.}\BibitemShut {Stop}%
\bibitem [{\citenamefont {Kamenev}(2011)}]{kamenev2011field}%
  \BibitemOpen
  \bibfield  {author} {\bibinfo {author} {\bibfnamefont {A.}~\bibnamefont
  {Kamenev}},\ }\href@noop {} {\emph {\bibinfo {title} {Field theory of
  non-equilibrium systems}}}\ (\bibinfo  {publisher} {Cambridge University
  Press},\ \bibinfo {year} {2011})\BibitemShut {NoStop}%
\bibitem [{sm()}]{sm}%
  \BibitemOpen
  \href@noop {} {}\bibinfo {note} {See supplementary material for: (1) The
  explicit form of the collision integral in the generalized Boltzmann
  equation, along with the generalized distribution function matrix for the
  probe and bath. (2) The equivalence between the Complex SYK (CSYK) model with
  the quasi-particle approximation and the Brownian Complex SYK (BCSYK) model.
  (3) The derivation of the entropy dynamics for the BCSYK(3,1) model using
  scramblon effective theory.}\BibitemShut {Stop}%
\bibitem [{\citenamefont {Islam}\ \emph {et~al.}(2015)\citenamefont {Islam},
  \citenamefont {Ma}, \citenamefont {Preiss}, \citenamefont {Tai},
  \citenamefont {Lukin}, \citenamefont {Rispoli},\ and\ \citenamefont
  {Greiner}}]{Islam:2015mom}%
  \BibitemOpen
  \bibfield  {author} {\bibinfo {author} {\bibfnamefont {R.}~\bibnamefont
  {Islam}}, \bibinfo {author} {\bibfnamefont {R.}~\bibnamefont {Ma}}, \bibinfo
  {author} {\bibfnamefont {P.~M.}\ \bibnamefont {Preiss}}, \bibinfo {author}
  {\bibfnamefont {M.~E.}\ \bibnamefont {Tai}}, \bibinfo {author} {\bibfnamefont
  {A.}~\bibnamefont {Lukin}}, \bibinfo {author} {\bibfnamefont
  {M.}~\bibnamefont {Rispoli}},\ and\ \bibinfo {author} {\bibfnamefont
  {M.}~\bibnamefont {Greiner}},\ }\bibfield  {title} {\bibinfo {title}
  {{Measuring entanglement entropy through the interference of quantum
  many-body twins}}\ }\href {https://doi.org/10.1038/nature15750}
  {10.1038/nature15750} (\bibinfo {year} {2015}),\ \Eprint
  {https://arxiv.org/abs/1509.01160} {arXiv:1509.01160 [cond-mat.quant-gas]}
  \BibitemShut {NoStop}%
\bibitem [{\citenamefont {Kaufman}\ \emph {et~al.}(2016)\citenamefont
  {Kaufman}, \citenamefont {Tai}, \citenamefont {Lukin}, \citenamefont
  {Rispoli}, \citenamefont {Schittko}, \citenamefont {Preiss},\ and\
  \citenamefont {Greiner}}]{doi:10.1126/science.aaf6725}%
  \BibitemOpen
  \bibfield  {author} {\bibinfo {author} {\bibfnamefont {A.~M.}\ \bibnamefont
  {Kaufman}}, \bibinfo {author} {\bibfnamefont {M.~E.}\ \bibnamefont {Tai}},
  \bibinfo {author} {\bibfnamefont {A.}~\bibnamefont {Lukin}}, \bibinfo
  {author} {\bibfnamefont {M.}~\bibnamefont {Rispoli}}, \bibinfo {author}
  {\bibfnamefont {R.}~\bibnamefont {Schittko}}, \bibinfo {author}
  {\bibfnamefont {P.~M.}\ \bibnamefont {Preiss}},\ and\ \bibinfo {author}
  {\bibfnamefont {M.}~\bibnamefont {Greiner}},\ }\bibfield  {title} {\bibinfo
  {title} {Quantum thermalization through entanglement in an isolated many-body
  system},\ }\href {https://doi.org/10.1126/science.aaf6725} {\bibfield
  {journal} {\bibinfo  {journal} {Science}\ }\textbf {\bibinfo {volume}
  {353}},\ \bibinfo {pages} {794} (\bibinfo {year} {2016})},\ \Eprint
  {https://arxiv.org/abs/https://www.science.org/doi/pdf/10.1126/science.aaf6725}
  {https://www.science.org/doi/pdf/10.1126/science.aaf6725} \BibitemShut
  {NoStop}%
\bibitem [{\citenamefont {Brydges}\ \emph {et~al.}(2019)\citenamefont
  {Brydges}, \citenamefont {Elben}, \citenamefont {Jurcevic}, \citenamefont
  {Vermersch}, \citenamefont {Maier}, \citenamefont {Lanyon}, \citenamefont
  {Zoller}, \citenamefont {Blatt},\ and\ \citenamefont
  {Roos}}]{Brydges:2019wut}%
  \BibitemOpen
  \bibfield  {author} {\bibinfo {author} {\bibfnamefont {T.}~\bibnamefont
  {Brydges}}, \bibinfo {author} {\bibfnamefont {A.}~\bibnamefont {Elben}},
  \bibinfo {author} {\bibfnamefont {P.}~\bibnamefont {Jurcevic}}, \bibinfo
  {author} {\bibfnamefont {B.}~\bibnamefont {Vermersch}}, \bibinfo {author}
  {\bibfnamefont {C.}~\bibnamefont {Maier}}, \bibinfo {author} {\bibfnamefont
  {B.~P.}\ \bibnamefont {Lanyon}}, \bibinfo {author} {\bibfnamefont
  {P.}~\bibnamefont {Zoller}}, \bibinfo {author} {\bibfnamefont
  {R.}~\bibnamefont {Blatt}},\ and\ \bibinfo {author} {\bibfnamefont {C.~F.}\
  \bibnamefont {Roos}},\ }\bibfield  {title} {\bibinfo {title} {{Probing
  R\'enyi entanglement entropy via randomized measurements}},\ }\href
  {https://doi.org/10.1126/science.aau4963} {\bibfield  {journal} {\bibinfo
  {journal} {Science}\ }\textbf {\bibinfo {volume} {364}},\ \bibinfo {pages}
  {aau4963} (\bibinfo {year} {2019})}\BibitemShut {NoStop}%
\bibitem [{\citenamefont {Elben}\ \emph {et~al.}(2020)\citenamefont {Elben}
  \emph {et~al.}}]{Elben:2020hpu}%
  \BibitemOpen
  \bibfield  {author} {\bibinfo {author} {\bibfnamefont {A.}~\bibnamefont
  {Elben}} \emph {et~al.},\ }\bibfield  {title} {\bibinfo {title} {{Mixed-state
  entanglement from local randomized measurements}},\ }\href
  {https://doi.org/10.1103/PhysRevLett.125.200501} {\bibfield  {journal}
  {\bibinfo  {journal} {Phys. Rev. Lett.}\ }\textbf {\bibinfo {volume} {125}},\
  \bibinfo {pages} {200501} (\bibinfo {year} {2020})},\ \Eprint
  {https://arxiv.org/abs/2007.06305} {arXiv:2007.06305 [quant-ph]} \BibitemShut
  {NoStop}%
\bibitem [{\citenamefont {{van Enk}}\ and\ \citenamefont
  {{Beenakker}}(2012)}]{2012PhRvL.108k0503V}%
  \BibitemOpen
  \bibfield  {author} {\bibinfo {author} {\bibfnamefont {S.~J.}\ \bibnamefont
  {{van Enk}}}\ and\ \bibinfo {author} {\bibfnamefont {C.~W.~J.}\ \bibnamefont
  {{Beenakker}}},\ }\bibfield  {title} {\bibinfo {title} {{Measuring
  Tr{\ensuremath{\rho}}$^{n}$ on Single Copies of {\ensuremath{\rho}} Using
  Random Measurements}},\ }\href
  {https://doi.org/10.1103/PhysRevLett.108.110503} {\bibfield  {journal}
  {\bibinfo  {journal} {\prl}\ }\textbf {\bibinfo {volume} {108}},\ \bibinfo
  {eid} {110503} (\bibinfo {year} {2012})},\ \Eprint
  {https://arxiv.org/abs/1112.1027} {arXiv:1112.1027 [quant-ph]} \BibitemShut
  {NoStop}%
\bibitem [{\citenamefont {{Elben}}\ \emph {et~al.}(2018)\citenamefont
  {{Elben}}, \citenamefont {{Vermersch}}, \citenamefont {{Dalmonte}},
  \citenamefont {{Cirac}},\ and\ \citenamefont
  {{Zoller}}}]{2018PhRvL.120e0406E}%
  \BibitemOpen
  \bibfield  {author} {\bibinfo {author} {\bibfnamefont {A.}~\bibnamefont
  {{Elben}}}, \bibinfo {author} {\bibfnamefont {B.}~\bibnamefont
  {{Vermersch}}}, \bibinfo {author} {\bibfnamefont {M.}~\bibnamefont
  {{Dalmonte}}}, \bibinfo {author} {\bibfnamefont {J.~I.}\ \bibnamefont
  {{Cirac}}},\ and\ \bibinfo {author} {\bibfnamefont {P.}~\bibnamefont
  {{Zoller}}},\ }\bibfield  {title} {\bibinfo {title} {{R{\'e}nyi Entropies
  from Random Quenches in Atomic Hubbard and Spin Models}},\ }\href
  {https://doi.org/10.1103/PhysRevLett.120.050406} {\bibfield  {journal}
  {\bibinfo  {journal} {\prl}\ }\textbf {\bibinfo {volume} {120}},\ \bibinfo
  {eid} {050406} (\bibinfo {year} {2018})},\ \Eprint
  {https://arxiv.org/abs/1709.05060} {arXiv:1709.05060 [quant-ph]} \BibitemShut
  {NoStop}%
\bibitem [{\citenamefont {{Elben}}\ \emph {et~al.}(2019)\citenamefont
  {{Elben}}, \citenamefont {{Vermersch}}, \citenamefont {{Roos}},\ and\
  \citenamefont {{Zoller}}}]{2019PhRvA..99e2323E}%
  \BibitemOpen
  \bibfield  {author} {\bibinfo {author} {\bibfnamefont {A.}~\bibnamefont
  {{Elben}}}, \bibinfo {author} {\bibfnamefont {B.}~\bibnamefont
  {{Vermersch}}}, \bibinfo {author} {\bibfnamefont {C.~F.}\ \bibnamefont
  {{Roos}}},\ and\ \bibinfo {author} {\bibfnamefont {P.}~\bibnamefont
  {{Zoller}}},\ }\bibfield  {title} {\bibinfo {title} {{Statistical
  correlations between locally randomized measurements: A toolbox for probing
  entanglement in many-body quantum states}},\ }\href
  {https://doi.org/10.1103/PhysRevA.99.052323} {\bibfield  {journal} {\bibinfo
  {journal} {\pra}\ }\textbf {\bibinfo {volume} {99}},\ \bibinfo {eid} {052323}
  (\bibinfo {year} {2019})},\ \Eprint {https://arxiv.org/abs/1812.02624}
  {arXiv:1812.02624 [quant-ph]} \BibitemShut {NoStop}%
\end{thebibliography}%

\ifarXiv


\section*{Supplementary material}

# Supplementary Material for Scrambling Enabled Entropy Accumulation in Open Quantum Systems"

Yuke Zhang,[1] Zeyu Liu,[1] Shuo Zhang,[1] Langxuan Chen,[2] and Pengfei Zhang[1,3,4,5,*]

[1]*Department of Physics, Fudan University, Shanghai, 200438, China*
[2]*School of Physics, Xi'an Jiaotong University, Xi'an 710049, China*
[3]*State Key Laboratory of Surface Physics, Fudan University, Shanghai, 200438, China*
[4]*Shanghai Qi Zhi Institute, AI Tower, Xuhui District, Shanghai 200232, China*
[5]*Hefei National Laboratory, Hefei 230088, China*
(Dated: February 11, 2025)

In this supplementary material, we present: (1) The explicit form of the collision integral in the generalized Boltzmann equation, along with the generalized distribution function matrix for the probe and bath. (2) The equivalence between the Complex SYK (CSYK) model with the quasi-particle approximation and the Brownian Complex SYK (BCSYK) model. (3) The derivation of the entropy dynamics for the BCSYK(3,1) model using scramblon effective theory.

## 1. GENERALIZED BOLTZMANN EQUATION

We begin with the Schwinger-Dyson equation. After introducing the center-of-mass time $t$ and the relative time $t_r$, we obtain:

$$\frac{\partial G_{S,\mathbf{k}}^{ss'}(t,t)}{\partial t} = \int_{-\infty}^{+\infty} dt_r \left[ (-1)^s \Sigma_{S,\mathbf{k}}^{ss''}\left(t+\frac{t_r}{2}, t-\frac{t_r}{2}\right) G_{S,\mathbf{k}}^{s''s'}\left(t-\frac{t_r}{2}, t+\frac{t_r}{2}\right) \right. \\ \left. - (-1)^{s'} G_{S,\mathbf{k}}^{ss''}\left(t+\frac{t_r}{2}, t-\frac{t_r}{2}\right) \Sigma_{S,\mathbf{k}}^{s''s'}\left(t-\frac{t_r}{2}, t+\frac{t_r}{2}\right) \right]. \tag{1}$$

The self-energy is given by melon diagrams:

$$\Sigma_{S,\mathbf{k}}^{ss'}(t_1, t_2) = -P^{ss'} \int \frac{dk_1^D}{(2\pi)^D} \frac{dk_2^D}{(2\pi)^D} \left[ J^2 G_{S,\mathbf{k}_2}^{ss'}(t_1, t_2) G_{S,\mathbf{k}_1+\mathbf{k}-\mathbf{k}_2}^{ss'}(t_1, t_2) G_{S,\mathbf{k}_1}^{s's}(t_2, t_1) \right. \\ \left. + \kappa^2 G_{P,\mathbf{k}_2}^{ss'}(t_1, t_2) G_{P,\mathbf{k}_1+\mathbf{k}-\mathbf{k}_2}^{ss'}(t_1, t_2) G_{P,\mathbf{k}_1}^{s's}(t_2, t_1) + V^2 G_{B,\mathbf{k}_2}^{ss'}(t_1, t_2) G_{B,\mathbf{k}_1+\mathbf{k}-\mathbf{k}_2}^{ss'}(t_1, t_2) G_{B,\mathbf{k}_1}^{s's}(t_2, t_1) \right] \tag{2}$$

$P^{ss'} = -1$ if both $s$ and $s'$ are $u$ or $d$ contours. Otherwise, $P^{ss'} = 1$.
By assumption, $M, \tilde{M} \gg N \gg 1$, we ignore the correction to the Green's function of the probe and bath that comes from the system fermions. Similar to the system Green's function, for $t_1 > t_2$, we also parameterize the probe and bath generalized distribution function matrix as: $G_{P/B,\mathbf{k}}^{ss'}(t_1, t_2) = e^{-i\epsilon_{P/B,\mathbf{k}} t_{12}} F_{P/B,\mathbf{k}}^{ss'}\left(\frac{t_1+t_2}{2}\right)$.

$$F_{P,\mathbf{k}}^{ss'}(t) = \begin{pmatrix} -n_{2\beta} & -w_{2\beta} & -w_{2\beta} & n_{2\beta}-1 \\ w_{2\beta} & 1-n_{2\beta} & -n_{2\beta} & -w_{2\beta} \\ w_{2\beta} & 1-n_{2\beta} & -n_{2\beta} & -w_{2\beta} \\ n_{2\beta} & w_{2\beta} & w_{2\beta} & 1-n_{2\beta} \end{pmatrix}_{ss'}, \quad F_{B,\mathbf{k}}^{ss'}(t) = \begin{pmatrix} -n_\beta & n_\beta-1 & 0 & 0 \\ n_\beta & 1-n_\beta & 0 & 0 \\ 0 & 0 & -n_\beta & n_\beta-1 \\ 0 & 0 & n_\beta & 1-n_\beta \end{pmatrix}_{ss'}. \tag{3}$$

The expression for $G_{P/B,\mathbf{k}}^{ss'}(t_1, t_2)$ with $t_1 < t_2$ is obtained by replacing $F_{P/B,\mathbf{k}}^{ss'}$ with $F_{P/B,\mathbf{k}}^{ss'} - (-1)^s \delta_{ss'}$, where $(-1)^s = -1$ for the $d$ branch and 1 for the $u$ branch.
The evolution of $f_{z,\mathbf{k}}$ can be derived using the Schwinger-Dyson equation:

$$\frac{\partial f_{z,\mathbf{k}}(t)}{\partial t} = J^2 \int d\mathcal{M}_S \, \text{St}_{S,z}[f_{z,\mathbf{k}}] + \kappa^2 \int d\mathcal{M}_P \, \text{St}_{P,z}[f_{z,\mathbf{k}}] + V^2 \int d\mathcal{M}_B \, \text{St}_{B,z}[f_{z,\mathbf{k}}] \tag{4}$$

$\text{St}_z[f]$ are complex combinations of distribution functions:

$$\begin{aligned}
\text{St}_{S,0}[f] &= f_1(k_2)f_1(k_3)f_2(k)f_2(k_1) - f_1(k)f_1(k_1)f_2(k_2)f_2(k_3) \\
&\quad + f_4(k_2)f_4(k_3)f_5(k)f_5(k_1) - f_4(k)f_4(k_1)f_5(k_2)f_5(k_3) \\
\text{St}_{S,1}[f] &= f_1(k)\left[f_0(k_1)\left[f_0(k_2)(1-2f_0(k_3)) + f_0(k_3) - 1\right] + f_0(k_2)f_0(k_3)\right] \\
&\quad + \left[1 - 2f_0(k)\right]f_1(k_2)f_1(k_3)f_2(k_1) \\
&\quad - 2f_3(k)f_4(k_2)f_4(k_3)f_5(k_1) + 2f_3(k_1)f_3(k_2)f_3(k_3)f_4(k) \\
\text{St}_{S,2}[f] &= -f_2(t)\left[f_0(t)\left[f_0(t)(1-2f_0(t)) + f_0(t) - 1\right] + f_0(t)f_0(t)\right] \\
&\quad + \left[2f_0(t) - 1\right]f_1(t)f_2(t)f_2(t) \\
&\quad - 2f_3(t)f_4(t)f_5(t)f_5(t) + 2f_3(t)f_3(t)f_3(t)f_5(t) \\
\text{St}_{S,3}[f] &= f_1(t)f_2(t)f_2(t)f_4(t) + f_1(t)f_1(t)f_2(t)f_5(t) \\
&\quad - f_1(t)f_4(t)f_5(t)f_5(t) - f_2(t)f_4(t)f_4(t)f_5(t) \\
\text{St}_{S,4}[f] &= f_4(t)\left(f_0(t)(-2f_0(t)f_0(t) + f_0(t) + f_0(t) - 1) + f_0(t)f_0(t)\right) \\
&\quad + (1 - 2f_0(t))f_4(t)f_4(t)f_5(t) \\
&\quad + 2f_1(t)f_1(t)f_2(t)f_3(t) - 2f_1(t)f_3(t)f_3(t)f_3(t) \\
\text{St}_{S,5}[f] &= f_5(t)\left[f_0(t)(2f_0(t)f_0(t) - f_0(t) - f_0(t) + 1) - f_0(t)f_0(t)\right] \\
&\quad + \left[2f_0(t) - 1\right]f_4(t)f_5(t)f_5(t) \\
&\quad + 2f_1(t)f_2(t)f_2(t)f_3(t) - 2f_2(t)f_3(t)f_3(t)f_3(t)
\end{aligned} \quad (5)$$

$\text{St}_{P,z}[f]$ can be obtained by replacing $f_3(k_i) = f_4(k_i) = f_5(k_i) = 0$ and $f_0(k_i) = f_1(k_i) = 1 - f_2(k_i) = n_\beta$. $\text{St}_{B,z}[f]$ can be obtained by replacing $f_1(k_i) = f_2(k_i) = f_3(k_i) = w_{2\beta}$ and $f_0(k_i) = f_4(k_i) = 1 - f_5(k_i) = n_{2\beta}$.

## 2. PROOF OF THE EQUIVALENCE

We first introduce the Brownian Complex SYK model, which is restricted to a four-point coupling form:

$$\begin{aligned}
H_C(t) &= {\sum_{\{o_l\}}}' \sum_{\{i_l, j_l, a_l, b_l\}} J_{i_l, j_l, a_l, b_l}(t)\, c^\dagger_{S,i_1} \cdots c^\dagger_{S,_{o_1}} c_{S,j_1} \cdots c_{S,j_{o_2}} c^\dagger_{B,a_1} \cdots c^\dagger_{B,a_{o_3}} c_{B,b_1} \cdots c_{B,b_{o_4}} \\
&\quad + {\sum_{\{p_l\}}}' \sum_{\{i_l, j_l, a_l, b_l\}} \kappa_{i_l, j_l, a_l, b_l}(t)\, c^\dagger_{S,i_1} \cdots c^\dagger_{S,_{p_1}} c_{S,j_1} \cdots c_{S,j_{p_2}} c^\dagger_{P,a_1} \cdots c^\dagger_{P,a_{p_3}} c_{P,b_1} \cdots c_{P,b_{p_4}} \\
&\quad + {\sum_{\{q_l\}}}' \sum_{\{i_l, j_l, a_l, b_l\}} V_{i_l, j_l, a_l, b_l}(t)\, c^\dagger_{S,i_1} \cdots c^\dagger_{S,_{q_1}} c_{S,j_1} \cdots c_{S,j_{q_2}} c^\dagger_{B,a_1} \cdots c^\dagger_{B,a_{q_3}} c_{B,b_1} \cdots c_{B,b_{q_4}}
\end{aligned} \quad (6)$$

Where $o_1 + o_2 + o_3 + o_4 = p_1 + p_2 + p_3 + p_4 = q_1 + q_2 + q_3 + q_4 = 4$, and $J_{i_l,j_l,a_l,b_l}(t)$, $\kappa_{i_l,j_l,a_l,b_l}(t)$, $V_{i_l,j_l,a_l,b_l}(t)$ are Brownian variables.

$$\overline{J_{i_l,j_l,a_l,b_l}(t) J_{i_l,j_l,a_l,b_l}(t')^*} = \frac{o_2 J_{\{o_l\}} \delta(t-t')}{\prod_{l=1}^{4} o_l! N^{o_1+o_2-1} M^{o_3+o_4}}. \quad \overline{V_{i_l,j_l,a_l,b_l}(t) V_{i_l,j_l,a_l,b_l}(t')^*} = \frac{p_2 V_{\{p_l\}} \delta(t-t')}{\prod_{l=1}^{4} p_l! N^{p_1+p_2-1} M^{p_3+p_4}}. \quad (7)$$

$$\overline{\kappa_{i_l,j_l,a_l,b_l}(t) \kappa_{i_l,j_l,a_l,b_l}(t')^*} = \frac{q_2 \kappa_{\{q_l\}} \delta(t-t')}{\prod_{l=1}^{4} q_l! N^{q_1+q_2-1} \tilde{M}^{q_3+q_4}}.$$

By choosing $\mathbf{o} = (2, 2, 0, 0)$, $\mathbf{p} = (1, 0, 1, 2)$, and $\mathbf{q} = (1, 0, 1, 2)$, we construct the BCSYK(4) model:

$$H(t) = \sum_{i<j,k<l} J_{ijkl}(t) c^\dagger_{S,i} c^\dagger_{S,j} c_{S,k} c_{S,l} + \sum_{i, m<p<q}\left[\kappa_{impq}(t) c^\dagger_{S,i} c^\dagger_{P,m} c_{P,p} c_{P,q} + \text{H.C.}\right] + \sum_{i, m<p<q}\left[V_{impq}(t) c^\dagger_{S,i} c^\dagger_{B,m} c_{B,p} c_{B,q} + \text{H.C.}\right] \quad (8)$$

Similarly, the self-energy is given by melon diagrams:

$$\begin{aligned}
\Sigma^{ss'}_S(t_1, t_2) &= -P^{ss'}\left[J G^{s,s'}_S(t_1, t_2)^2 G^{s',s}_S(t_2, t_1) + \kappa G^{s,s'}_P(t_1, t_2)^2 G^{s',s}_P(t_2, t_1) + V G^{s,s'}_B(t_1, t_2)^2 G^{s',s}_B(t_2, t_1)\right]\delta(t_1 - t_2) \\
&= \tilde{\Sigma}^{ss'}_S(t_1, t_2)\delta(t_1 - t_2)
\end{aligned} \quad (9)$$

Substituting Eq.(9) into Eq.(1), we obtain:

$$\frac{\partial G_S^{ss'}(t,t)}{\partial t} = (-1)^s \tilde{\Sigma}_S^{ss''}(t,t) G_S^{s''s'}(t,t) - (-1)^{s'} G_S^{ss''}(t,t) \tilde{\Sigma}_S^{s''s'}(t,t) \tag{10}$$

The relative time dependence vanishes after we integrate the delta function. After identifying $G_S^{ss'}(t)$ with $F_S^{ss'}(t)$, the final result coincides with the Boltzmann equation for the flat band system, where we use the quasi-particle approximation to parameterize the relative-time contribution. We also notice that the substitution rule is $J \longleftrightarrow \tilde{J}$, $\kappa \longleftrightarrow \tilde{\kappa}$, and $V \longleftrightarrow \tilde{V}$.

## 3. ENTROPY DYNAMICS FOR SOLVABLE MODEL

By choosing $\mathbf{o} = (2, 1, 0, 1)$, $\mathbf{p} = (1, 0, 1, 2)$, and $\mathbf{q} = (1, 0, 1, 2)$, we construct the BCSYK(3,1) model:

$$\begin{aligned}
H(t) &= \sum_{i<j,k,m} \left[ J_{ijkm}(t) c_{S,i}^\dagger c_{S,j}^\dagger c_{S,k} c_{V,m} + \text{H.C.} \right] + \sum_{i,m<p<q} \left[ \kappa_{impq}(t) c_{S,i}^\dagger c_{P,m}^\dagger c_{P,p} c_{P,q} + \text{H.C.} \right] \\
&+ \sum_{i,m<p<q} \left[ V_{impq}(t) c_{S,i}^\dagger c_{B,m}^\dagger c_{B,p} c_{B,q} + \text{H.C.} \right]
\end{aligned} \tag{11}$$

Unlike the BCSYK(4) model, the entropy dynamics of the BCSYK(3,1) model can be solved exactly. In the limit of $\kappa \ll V, J$, ②$(t)$ is the only non-trivial part, which is closely related to the OTOC. The Scramblon Effective Theory captures the chaotic signature of the OTOC and has been used to study entropy dynamics in bosonic closed systems [1, 2]. We generalize the formalism to the BCSYK(3,1) model and calculate the entropy dynamics. We begin by taking the path integral representation of the entropy contour.

$$e^{-S^{(2)}(t,0)} = \text{tr}_{SB} \left( \text{tr}_P \, e^{-iHt} \rho_0 e^{iHt} \right)^2 = \tag{12}$$

Where $\rho_0$ is the initial Gibbs state $\frac{e^{-\mu \hat{N}}}{Z}$, and $\hat{N}$ is the total particle number operator. We integrate out the probe fermions and obtain a source term:

$$\begin{aligned}
S_{int} &= \sum_{i=1}^N \int dt_1 dt_2 \Big[ \bar{\psi}_{S,1}^i(t_1) \Sigma_{12}(t_1,t_2) \psi_{S,2}^i(t_2) + \bar{\psi}_{S,2}^i(t_1) \Sigma_{21}(t_1,t_2) \psi_{S,1}^i(t_2) \\
&\quad + \bar{\psi}_{S,3}^i(t_1) \Sigma_{34}(t_1,t_2) \psi_{S,4}^i(t_2) + \bar{\psi}_{S,4}^i(t_1) \Sigma_{43}(t_1,t_2) \psi_{S,3}^i(t_2) \Big] \\
&= -\kappa \sum_{i=1}^N \int dt \Big[ (-1+n_\mu)^2 n_\mu \bar{\psi}_{S,1}^i(t) \psi_{S,2}^i(t) + n_\mu^2(-1+n_\mu) \bar{\psi}_{S,2}^i(t) \psi_{S,1}^i(t) \\
&\quad + (-1+n_\mu)^2 n_\mu \bar{\psi}_{S,3}^i(t) \psi_{S,4}^i(t) + n_\mu^2(-1+n_\mu) \bar{\psi}_{S,4}^i(t) \psi_{S,3}^i(t) \Big]
\end{aligned} \tag{13}$$

$n_\mu = \frac{1}{e^\mu+1}$ is the average occupation number.

$$②(t) = \text{[diagrams]} \quad (14)$$

The blue dashed line represents the interaction between the $u$ and $d$ contours, which comes from integrating out the probe fermions.

We identify the six diagrams above using the Scramblon Effective Theory.

$$\begin{aligned}
②(t) &= \int_0^\infty dy \left[h_\psi^A(y,-i\mu)_{2\mu} + [h_{\psi^\dagger}^A(y,-i\mu)_{2\mu}\right] e^{4\kappa N \int dt n_\mu(1-n_\mu)\left[f_\psi^R(y\frac{e^{i\varkappa(\frac{\mu}{2}-it)}}{C},-i\mu)_{2\mu}-G_\psi(-i\mu)\right]} \\
&+ \int_0^\infty dy \left[h_\psi^A(y,-i\mu)_{2\mu} + [h_{\psi^\dagger}^A(y,-i\mu)_{2\mu}\right] e^{4\kappa N \int dt n_\mu(1-n_\mu)\left[f_\psi^R(y\frac{e^{-i\varkappa(\frac{\mu}{2}+it)}}{C},-i\mu)_{2\mu}-G_\psi(-i\mu)\right]} \\
&- 2\int_0^\infty dy \left[h_\psi^A(y,-2i\mu)_{2\mu} + [h_{\psi^\dagger}^A(y,-2i\mu)_{2\mu}\right] e^{4\kappa N \int dt n_\mu(1-n_\mu)\left[f_\psi^R(y\frac{e^{\varkappa t}}{C},-i\mu)_{2\mu}-G_\psi(-i\mu)\right]} \\
&\approx 2\int_0^\infty dy h_\psi^A(y,-i\mu)_{2\mu}\, e^{-y\frac{4\kappa N}{\varkappa C}n_\mu(1-n_\mu)\Upsilon_\psi^R(-i\mu)_{2\mu}e^{i\varkappa \frac{\mu}{2}}(e^{\varkappa t}-1)} + 2\int_0^\infty dy h_\psi^A(y,-i\mu)_{2\mu}\, e^{-y\frac{4\kappa N}{\varkappa C}n_\mu(1-n_\mu)\Upsilon_\psi^R(-i\mu)_{2\mu}e^{-i\varkappa \frac{\mu}{2}}(e^{\varkappa t}-1)} \qquad (15)\\
&- 2\int_0^\infty dy \left[h_\psi^A(y,-2i\mu)_{2\mu} + h_{\psi^\dagger}^A(y,-2i\mu)_{2\mu}\right] e^{-y\frac{4\kappa N}{\varkappa C}n_\mu(1-n_\mu)\Upsilon_\psi^R(-i\mu)_{2\mu}(e^{\varkappa t}-1)} \\
&= 2f_\psi^A\left[\frac{4\kappa N}{\varkappa C}n_\mu(1-n_\mu)\Upsilon_\psi^R(-i\mu)_{2\mu}e^{i\varkappa \frac{\mu}{2}}(e^{\varkappa t}-1),-i\mu\right]_{2\mu} + 2f_\psi^A\left[\frac{4\kappa N}{\varkappa C}n_\mu(1-n_\mu)\Upsilon_\psi^R(-i\mu)_{2\mu}e^{-i\varkappa \frac{\mu}{2}}(e^{\varkappa t}-1),-i\mu\right]_{2\mu} \\
&- 2f_\psi^A\left[\frac{4\kappa N}{\varkappa C}n_\mu(1-n_\mu)\Upsilon_\psi^R(-i\mu)_{2\mu}(e^{\varkappa t}-1),-2i\mu\right]_{2\mu} - 2f_\psi^A\left[\frac{4\kappa N}{\varkappa C}n_\mu(1-n_\mu)\Upsilon_\psi^R(-i\mu)_{2\mu}(e^{\varkappa t}-1),0\right]_{2\mu}
\end{aligned}$$

Where $f_O^{R/A}(z,\theta)_{n\mu}$, $h_O^{R/A}(y,\theta)_{n\mu}$, $\Upsilon_O^{R/A}(\theta)_{n\mu}$ are some auxiliary functions in Scramblon Effective Theory, which satisfy:

$$f_O^{R/A}(z,\theta)_{n\mu} = \sum_{m=0}^\infty \frac{(-z)^m}{m!}\Upsilon_O^{R/A,m}(\theta)_{n\mu}, \qquad f_O^{R/A}(z,\theta)_{n\mu} = \int_0^\infty dy\, h_O^{R/A}(y,\theta)_{n\mu} e^{-zy} \qquad (16)$$

The auxiliary functions for similar model (with direct hopping coupling) have been calculated [3]. The generalization to the BCSYK(3,1) model is straightforward.

$$\begin{aligned}
C &= 2N(1-r)^2 n_{2\mu}(1-n_\mu), \quad \Upsilon^{R/A,0}(t-i\mu)_{2\mu} = \sqrt{n_{2\mu}(1-n_{2\mu})}e^{-\frac{\Gamma}{2}|t|} \\
\Upsilon^{R/A,k\neq 0}&(t-i\mu)_{2\mu} = k!(1-r)\sqrt{n_{2\mu}(1-n_{2\mu})}e^{-\frac{\Gamma+k\varkappa}{2}|t|} \\
f_\psi^{R/A}(z,t)_{2\mu} &= (1-n_{2\mu})f(z,t) = (1-n_{2\mu})e^{-\frac{\Gamma}{2}|t|}\left(r + \frac{1-r}{1+ze^{-\varkappa\frac{|t|}{2}}}\right) \\
f_\psi^{R/A}(z,t-i\mu)_{2\mu} &= \sqrt{n_{2\mu}(1-n_{2\mu})}f(z,t) = \sqrt{n_{2\mu}(1-n_{2\mu})}e^{-\frac{\Gamma}{2}|t|}\left(r + \frac{1-r}{1+ze^{-\varkappa\frac{|t|}{2}}}\right) \\
f_\psi^{R/A}(z,t-2i\mu)_{2\mu} &= n_{2\mu}f(z,t) = n_{2\mu}e^{-\frac{\Gamma}{2}|t|}\left(r + \frac{1-r}{1+ze^{-\varkappa\frac{|t|}{2}}}\right)
\end{aligned} \qquad (17)$$

Where $r = \frac{V}{2J}$, $\Gamma = n_{2\mu}\left(1 - n_{2\mu}\right)(2J + V)$ is the decay rate of quasi-particles, and $\varkappa = n_{2\mu}\left(1 - n_{2\mu}\right)(2J - V)$ is the quantum Lyapunov exponent. Substituting Eq.(17) into Eq.(15), and adding the entropy contribution of ①$(t)$, we obtain the final results for the entropy dynamics:

$$\delta S^{(2)}(t) = ①(t) + ②(t) = 2(1-r)\left(1 - 2\sqrt{n_{2\beta}(1-n_{2\beta})}\right) \\ - 2(1-r)\left[\frac{1}{1+g(t)} - 2\sqrt{n_{2\beta}(1-n_{2\beta})} \frac{1 + g(t)\cos\left(\frac{\beta}{2}\varkappa\right)}{1 + 2g(t)\cos\left(\frac{\beta}{2}\varkappa\right) + g(t)^2}\right] \tag{18}$$

Here we have used the substitution rule: $\mu \to \beta(\epsilon - \mu)$, and $g(t) = \frac{2\varkappa}{\varkappa(1-r)} n_\beta\left(1 - n_\beta\right)(e^{\varkappa t} - 1)$. We take the limit $t \to \infty$ to obtain the saturated solution:

$$\delta S^{(2)}_{sat} = \theta(2J - V)\, 2(1-r)\left(1 - 2\sqrt{n_{2\beta}(1-n_{2\beta})}\right) \tag{19}$$

---

[1] Y. Gu, A. Kitaev, and P. Zhang, A two-way approach to out-of-time-order correlators, JHEP **03**, 133, arXiv:2111.12007 [hep-th].
[2] P. Zhang, Perturbative Page curve induced by external impulse, JHEP **09**, 056, arXiv:2305.18329 [cond-mat.stat-mech].
[3] P. Zhang and Z. Yu, Environment-induced information scrambling transition with charge conservations, AAPPS Bull. **34**, 19 (2024), arXiv:2403.08622 [quant-ph].


\fi

\end{document}